\documentclass[aps,twocolumn,prx,preprintnumbers,showpacs,amsmath,amssymb,superscriptaddress,longbibliography]{revtex4-1}
\usepackage[pdftex, colorlinks, citecolor=blue]{hyperref}   
\usepackage{graphicx}
\usepackage{dcolumn}
\usepackage{threeparttable}
\usepackage{multirow}
\usepackage{booktabs}
\usepackage{txfonts}
\usepackage{xcolor}
\usepackage{bm}
\usepackage{amssymb}
\usepackage{amsmath}
\usepackage{latexsym}
\usepackage{epsfig}
\usepackage{amsbsy}
\usepackage{array}
\usepackage{tabularx}
\usepackage{esvect} 
\usepackage{extarrows}
\usepackage{float}
\usepackage[british]{babel}

\usepackage[utf8]{inputenc}

\begin{document}

\title{Efficient small-cell sampling for machine-learning potentials of multi-principal element alloys}

\author{Yan Liu}
\affiliation{Shenyang National Laboratory for Materials Science, Institute of Metal Research, Chinese Academy of Sciences, 110016 Shenyang, China}
\affiliation{School of Materials Science and Engineering, University of Science and Technology of China, 110016 Shenyang, China}

\author{Jiantao Wang}
\affiliation{Shenyang National Laboratory for Materials Science, Institute of Metal Research, Chinese Academy of Sciences, 110016 Shenyang, China}
\affiliation{School of Materials Science and Engineering, University of Science and Technology of China, 110016 Shenyang, China}

\author{Hongkun Deng}
\affiliation{Shenyang National Laboratory for Materials Science, Institute of Metal Research, Chinese Academy of Sciences, 110016 Shenyang, China}
\affiliation{School of Materials Science and Engineering, University of Science and Technology of China, 110016 Shenyang, China}

\author{Yan Sun}
\affiliation{Shenyang National Laboratory for Materials Science, Institute of Metal Research, Chinese Academy of Sciences, 110016 Shenyang, China}

\author{Xing-Qiu Chen}
\email{xingqiu.chen@imr.ac.cn}
\affiliation{Shenyang National Laboratory for Materials Science, Institute of Metal Research, Chinese Academy of Sciences, 110016 Shenyang, China}

\author{Peitao Liu}
\email{ptliu@imr.ac.cn}
\affiliation{Shenyang National Laboratory for Materials Science, Institute of Metal Research, Chinese Academy of Sciences, 110016 Shenyang, China}

\begin{abstract}
Multi-principal element alloys (MPEAs) exhibit exceptional properties
but face significant challenges in developing accurate machine-learning potentials (MLPs)
due to their vast compositional and configurational complexity.
Here, we introduce an efficient small-cell sampling (SCS) method,
which allows for generating diverse and representative training datasets for MPEAs using only small-cell structures with just one and two elements,
thereby bypassing the computational overhead of iterative active learning cycles and large-cell density functional theory calculations.
The efficacy of the method is carefully validated through principal component analysis, extrapolation grades evaluation,
and root-mean-square errors and physical properties assessment on the TiZrHfCuNi system.
Further demonstrations on TiZrVMo, CoCrFeMnNi, and AlTiZrNbHfTa systems
accurately reproduce complex phenomena including phase transitions, chemical orderings, and thermodynamic properties.
This work establishes an efficient one-shot protocol for constructing high-quality training datasets across multiple elements,
laying a solid foundation for developing universal MLPs for MPEAs.
\end{abstract}

\maketitle

\section{Introduction}\label{sec:introduction}

Developing high-performance materials capable of meeting the rapidly evolving demands of technology represents a major challenge in modern materials science.
Traditional alloy design---primarily based on one or two principal elements with minor additions of other elements to
enhance properties---has nearly reached its plateau through composition and microstructure optimization.
In contrast, multi-principal element alloys (MPEAs), first introduced by Cantor~\cite{Cantor2004} and Yeh~\cite{Yeh2004},
have revolutionized conventional alloy design, offering transformative breakthroughs and opportunities for materials science.
Compared to traditional alloys, MPEAs exhibit unique characteristics, including the thermodynamic high-entropy effect,
severe lattice distortion, sluggish diffusion kinetics, and the cocktail effect on properties~\cite{Yeh2013,Tsai2014,Hsu2024}.
These effects endow MPEAs with exceptional thermal stability and mechanical properties~\cite{Zhang2018,George2019,Wang2021,Rao2022,Ma2024,Sohail2025},
outstanding resistance to irradiation damage~\cite{Tianxin2021,Khan2024}, superior corrosion resistance~\cite{Shi2017}, and remarkable catalytic performance~\cite{Ren2023,Cai2025}.

Despite their potential, the rational design of MPEAs presents significant challenges for both experimental research and theoretical calculations
due to their immense and intrinsically coupled compositional, configurational, and microstructural design spaces~\cite{Miracle2017,Widom2018,George2019,Hart2021,Yan2025}.
Specifically, the compositional space formed by say five principal elements with just 100 atoms
would result in $100!/(\frac{100}{5}!)^5\approx10^{66}$ possible equiatomic configurations and a nearly unbounded number of non-equiatomic configurations.
Second, the configurational space is complicated by the presence of chemical short-range order (SRO)~\cite{chen2021direct},
where the arrangement of atoms critically influences key properties such as corrosion and wear resistance.
Lastly, the microstructural space, including lattice distortions, dislocations, and interface structures, ultimately determines the mechanical performance of the alloys.

While density functional theory (DFT) offers valuable atomic-level insights, its unfavorable computational scaling limits its practical simulations to smaller systems,
which are often insufficient for accurately capturing emergent SRO or defects~\cite{Zhao2021}.
In contrast, molecular dynamics (MD) and Monte Carlo (MC) simulations facilitate the exploration of extended time and length scales;
however, they are hampered by the lack of accurate and transferable interatomic potentials, particularly for multicomponent alloys.
Although effective Hamiltonians for alloys can be constructed using the widely adopted cluster expansion (CE) method~\cite{1984Sanchez, 1994Fontaine, vandeWalle2002, 2002van_de_Walle_ATAT,ZHU202354}, their computational feasibility is often constrained in multicomponent systems
due to the combinatorial explosion of interatomic interactions among different chemical elements~\cite{PhysRevB.80.165122_2009,PhysRevLett.116.105501_2016}.
The recently proposed effective pair interaction (EPI) model enhances the conventional CE method, demonstrating efficient and robust performance
in predicting the configurational energy of MPEAs~\cite{EPI_NC_Santodonato2018,Zhang_MaterialsDesign2020,Liu_CMS2021,EPI_Acta2025,Liu_arXiv2025}.
However, the EPI model is fundamentally an Ising-like framework that only considers effective pair interactions, neglecting high-order interactions.
Furthermore, similar to the CE method, it operates as an on-site energy model, which does not allow to calculate forces and
stresses---essential components for structure relaxation and MD simulations, particularly in MPEAs that exhibit significant lattice distortion.

Machine-learning potentials (MLPs) overcome the limitations of the CE method
by employing flexible, high-dimensional parameterizations based on high-fidelity DFT data,
which allows for a more accurate description of the potential energy surfaces of
complex systems~\cite{doi:10.1021/acs.chemrev.0c00868, doi:10.1021/acs.chemrev.1c00022,
doi:10.1021/acs.chemrev.0c01111, doi:10.1021/acs.chemrev.1c00021, doi:10.1021/acs.chemrev.0c01303,Wen_2022,Wang2024,Dong2024,Jinnouchi2025}.
In recent years, MLPs have achieved significant advancements, not only in applications tailored to specific materials
but also in the development of universal models~\cite{Takamoto2022,M3GNet-Chen2022,CHGNet-Deng2023,ALIGNN-FF-Choudhary2023,GNoME2023,MACE-MP-0-2024,MACE-OFF,XIE2024,
DPA2-2024,NEP16-Song2024,DPA3-2025,MatterSim-2024,Orb-2024,SevenNet-Park2024,PET-MAD-2025,NEP89-2025,EquiformerV2-OMAT24}.
These universal models are pre-trained on large and diverse materials datasets~\cite{EquiformerV2-OMAT24,MP_Jain2013,JARVIS_Choudhary2020,Alexandria_Schmidt2023},
offering strong out-of-the-box performance while remaining highly adaptable---requiring only minimal fine-tuning for new systems.
By drastically reducing the computational cost of deployment, they lower the barrier to adopting MLPs across materials science applications.
Once fine-tuned, they achieve high task-specific accuracy, enabling accurate property predictions, efficient molecular dynamics simulations, and accelerated discovery of novel materials.

However, these universal potentials have been developed for inorganic and organic materials
and their application to alloys---particularly MPEAs---remains limited by the lack of diverse and high-quality datasets.
The inherent compositional, configurational, and microstructural complexities of MPEAs pose a formidable challenge,
resulting in only a few of MLPs developed for MPEAs to date~\cite{Kostiuchenko2019,Balyakin_JPCM2020_VZrNbHfTa,
Li2020,Yin2021, Byggmastar_PRB2021_MoNbTaVW,Santos-Florez2023,10.1103/PhysRevMaterials.7.045802,Chang_2024,Wu_2024}.
Notably, the recent study of Song \emph{et al.} developed a general-purpose MLP for 16 elemental metals and their alloys~\cite{NEP16-Song2024},
which was later extended by Liang \emph{et al.} to inorganic and organic materials across 89 elements~\cite{NEP89-2025}.
Nevertheless, these potentials were trained using a manually-curated dataset based on expert physical and chemical intuition,
which may introduce imbalanced sampling and redundant structures.
Additionally, the use of large supercells in these datasets significantly increases computational demands.
Active learning methods~\cite{AL_Botu2015,AL_Alessandro2015,MTP2017,VASP2019,Kulichenko2023}
can iteratively expand training datasets with less human intervention,
but they often require multiple refinement cycles to adequately explore the relevant phase space.
These cycles are tedious and resource-intensive, involving repeatedly identifying new structures, computing their quantum-mechanical properties, and refitting the potential.
This raises a critical question: Can a diverse and high-quality training dataset for MPEAs be generated efficiently and at low cost,
minimizing human intervention and eliminating the need for iterative refinement?

Recent advances have highlighted the advantages of training MLPs on small cells over large
supercells~\cite{smallcell_Meziere2023,smallcell_Pickard2023,smallcell_JCTC_Luo2023, Neugebauer_PRB2023,Meng2025,smallcell_Neugebauer_npj2025}.
Remarkably, datasets composed solely of small cells can effectively represent large supercells,
yielding MLPs with comparable accuracy but at a significantly lower computational cost.
Moreover, since MLPs learn partitioned atomic energies rather than total energies,
small-cell training eliminates the energy degeneracy issues inherent in large-cell training
and provides richer atomic-energy-resolved information.
Despite these successes, small-cell training has so far been limited to systems with up to three elements,
leaving its applicability to MPEAs with multi-principal elements unresolved.

Notably, the recent work of Li \emph{et al.}~\cite{Li_2024} offers critical insights for MLP development in MPEAs.
Their work demonstrates that the generalization from ordered/non-equimolar/low-order alloys to disordered/equimolar/high-order alloys
is much easier than the reverse processes~\cite{Li_2024}.
This finding significantly simplifies the development of MLPs for MPEAs,
as training datasets can be constructed using simpler structures (e.g., low-order ordered phases, non-equimolar compositions).
Note that the generalization from ordered/non-equimolar alloys to disordered/equimolar alloys aligns with the small-cell training philosophy,
while the generalization from low-order to high-order alloys mirrors established strategies in phase diagram calculations~\cite{Saunders1998CALPHAD}.
Moreover, the work of Song \emph{et al.}~\cite{NEP16-Song2024} confirmed that
one- and two-component systems suffice to represent the full chemical space of multicomponent alloys.

\begin{figure}
	\centering
	\includegraphics[width=0.48\textwidth,trim = {0.0cm 0.0cm 0.0cm 0.0cm}, clip]{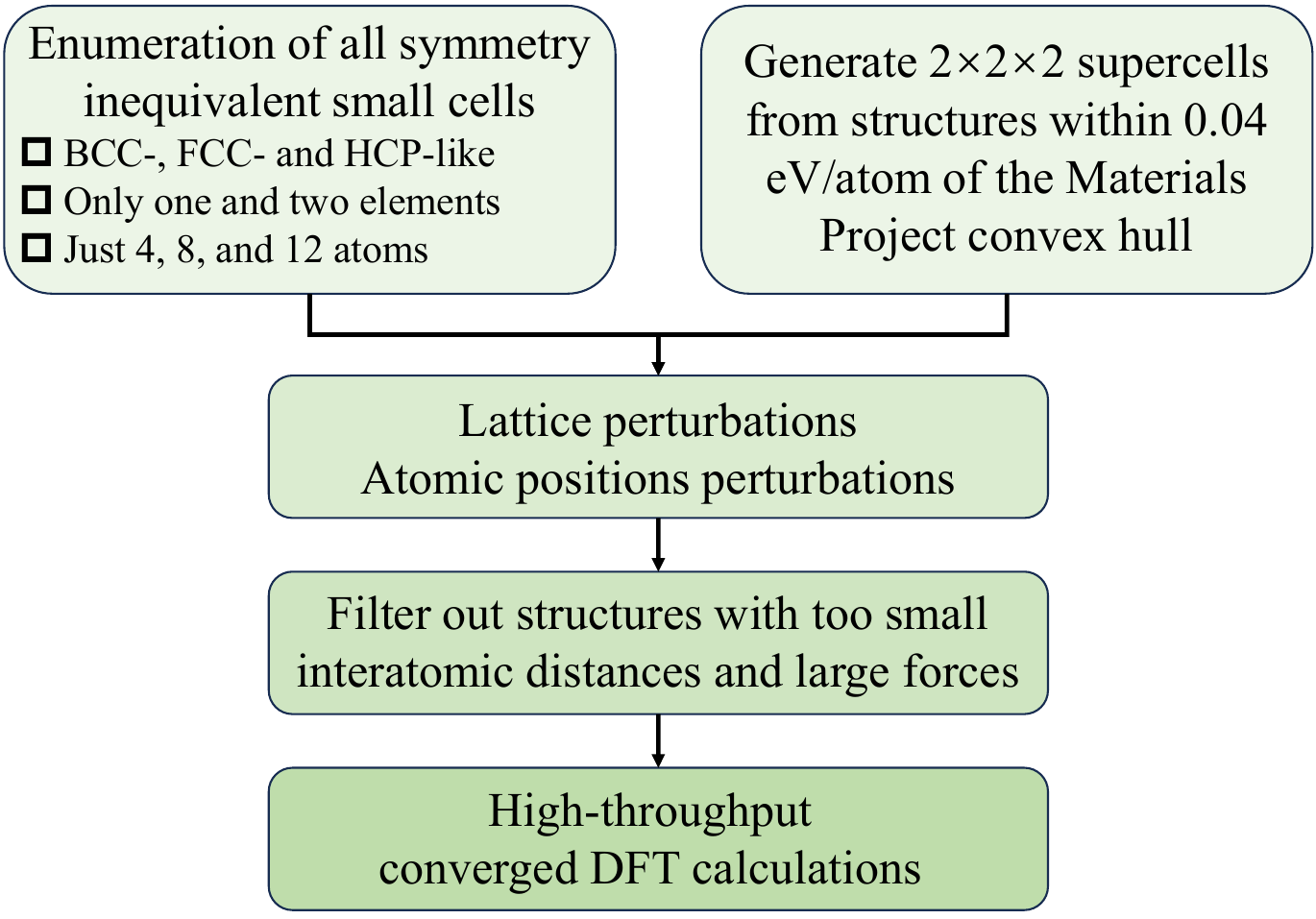}
\caption{Workflow illustrating the key steps in the SCS protocol.}
	\label{Fig1_workflow}
\end{figure}

Building on these insights, we introduce an efficient small-cell sampling (SCS) protocol for developing MLPs tailored to MPEAs (see Fig.~\ref{Fig1_workflow}).
Our method leverages the enumeration of all possible symmetry-inequivalent configurations with only one and two elements
within the BCC-, FCC-, and HCP-like small cells consisting of just 4, 8, and 12 atoms,
and selects structures within 0.04 eV/atom of the binary convex hull in the Materials Project (MP)~\cite{MP_Jain2013} database.
For the obtained structures, structural perturbations are then applied on lattice parameters and atomic positions to enhance phase-space sampling.
After filtering out structures with too small interatomic distances and large forces, high-throughput converged DFT calculations are conducted.
We show that this one-shot, low-component SCS strategy can generate a diverse and representative training dataset for multicomponent systems,
bypassing the computational overhead of iterative active learning cycles and large-cell DFT calculations.
We demonstrate the efficacy of this approach by investigating temperature-dependent SROs, energy convex hulls,
and phase transitions in four MPEAs, including TiZrHfCuNi, AlTiZrNbHfTa, CoCrFeMnNi, and TiZrVMo.
By introducing an efficient protocol for generating a diverse and high-quality training dataset across multiple elements,
this work lays a solid foundation for developing universal MLPs for MPEAs.

\section{Computational details}\label{sec:methods}

\subsection{First-principles calculations}\label{subsec:DFT_details}
All first-principles calculations were conducted using the Vienna \emph{ab initio} simulation package (VASP)~\cite{ PhysRevB.54.11169}.
The generalized gradient approximation of Perdew-Burke-Ernzerhof functional (PBE) was used for the exchange-correlation functional~\cite{10.1103/PhysRevLett.77.3865}.
The VASP recommended projector augmented wave (PAW) pseudopotentials were used~\cite{PhysRevB.59.1758}.
A plane wave cutoff of 600 eV and a $\Gamma$-centered $k$-point grid with a spacing of 0.2~$\AA^{-1}$ between $k$ points were employed.
The Gaussian smearing method with a smearing width of 0.05 eV was used.
The electronic optimization was performed until the total energy difference between two consecutive  iterations was less than 10$^{-6}$ eV.
The structures were optimized until the maximum forces below 0.01 eV/$\AA$.

\subsection{Training of machine-learning potential}\label{subsec:MACE_and_MTP}

In this work, we trained two different types of MLPs. The first is the physical descriptor-based linearly parameterized moment tensor potential (MTP)~\cite{Alexander2016},
which demonstrates the highest inference speed among descriptor-based MLPs~\cite{zuoPerformanceCostAssessment2020}.
The second is the higher-order equivariant message passing neural network (MACE)~\cite{Batatia2022mace},
characterized by its expressivity, data efficiency, and high accuracy~\cite{Batatia2022mace}.

For the MTP training, a cutoff radius of 6.0~$\AA$ was used, and the radial basis size was set to 8.
The maximum number of moment tensors involved in contraction was limited to 4.
The basis functions were optimized by refining the contraction process of moment tensors using our in-house code (IMR-MLP)~\cite{jiantao2024}.
The weights for energies, forces, and stress tensors were set to 1.0, 0.1, and 0.01, respectively.
The regression coefficients were obtained by a non-linear least square optimization using the Broyden-Fletcher-Goldfarb-Shanno algorithm.

The MACE models were constructed using the version 0.3.12~\cite{Batatia2022mace}.
The models employed a cutoff radius of 6~$\AA$ for atomic environments with 128 channels and $L = 2$.
The full dataset was partitioned into training and validation subsets with a 9:1 ratio.
The models were optimized using AMSGrad with a learning rate of 0.01 and a batch size of 40.
Model performance was assessed through a weighted root-mean-square-error (RMSE) loss function for the predictions of energies, forces, and stress tensors.
Initial weights were set to 1 for energies, 100 for forces, and 100 for stress tensors.
During the last 25\% of training epochs, the energy weight was progressively increased by a factor of 10 through three incremental steps.
To compare with the MTP, the final weights assigned to energies, forces, and stress tensors were aligned with those used by the MTP, specifically set at 1.0, 0.1, and 0.01, respectively.
The optimization protocol incorporated an early stopping criterion with a patience of 150 epochs.

\section{Results and discussion}\label{sec:Results and discussion}

\subsection{Outline of the SCS protocol}\label{subsec:Configuration Generation Protocol}

Figure~\ref{Fig1_workflow} outlines the key steps of the SCS protocol.
The first step involves constructing an initial dataset of small-cell structures.
We began by identifying all possible symmetry-inequivalent configurations
containing just one or two elements within BCC-, FCC-, and HCP-like small cells, specifically those comprising 4, 8, and 12 atoms.
This approach is grounded in several considerations:
($i$) BCC, FCC, and HCP atomic environments are predominant in MPEAs,
($ii$) one- and two-element systems can effectively represent the chemical space of multicomponent alloys~\cite{Li_2024,NEP16-Song2024},
and ($iii$) small conventional cells with up to 12 atoms can well capture the first-shell atomic environments characteristic of BCC, FCC, and HCP lattices.
The enumeration process was carried out using the automated simulation environment (ASE) python library~\cite{HjorthLarsen2017}
in conjunction with the derivative structure generation algorithm~\cite{10.1103/PhysRevB.77.224115}.
The number of enumerated symmetry-inequivalent configurations for the binary system is presented in Table~\ref{tab:binary_structure_distribution}.
In total, 635 small cells were obtained for each binary configuration.
To ensure the accuracy of thermodynamic convex hulls in phase diagrams,
it is essential to include not only solid solution phases but also ordered intermetallic compounds in the dataset.
To this end, we selected structures from the MP database that lie within 0.04 eV/atom of binary convex hulls.
The choice of this energy criterion is motivated by the fact that the accuracy of most universal MLPs falls within this range.
These selected ordered intermetallic compounds were then enlarged into 2$\times$2$\times$2 supercells,
a step we found critical for obtaining accurate phonons.

\begin{table}
	\centering
\caption{Enumeration of symmetry-inequivalent configurations for binary systems.}
	\begin{ruledtabular}
	\begin{tabular}{ccccc}
		& BCC & FCC & HCP & Sum \\
		\hline
		4-atom cell & 4 & 3 & 5 & 12 \\
		8-atom cell & 20 & 25 & 32 & 77 \\
		12-atom cell & 156 & 155 & 235 & 546 \\
		Sum & 180 & 183 & 272 & 635 \\
	\end{tabular}
\end{ruledtabular}
	\label{tab:binary_structure_distribution}
\end{table}

To further enhance phase-space sampling, for the structures obtained in the initial step,
we applied structural perturbations to both lattice parameters and atomic positions.
Specifically, we applied uniform scaling factors of 0.95, 1.00, and 1.05 on the lattice parameters of each structure.
This was followed by Gaussian-distributed random displacements of the atomic positions, with a standard deviation of 0.15~$\AA$.
This resulted in 1905 small structures for each binary configuration.
For unary systems, we implemented three additional perturbations to the atomic positions, yielding a total of 108 structures for each unary configuration.
We note that while it is possible to apply further structural perturbations to expand the dataset,
we aimed to balance dataset size with computational efficiency.
Our goal was to minimize computational costs while still adequately capturing the relevant phase space.
Afterwards, we filtered out structures with too small interatomic distances.
Specifically, any structure where the distance between two atoms was less than the sum of their PAW sphere radii was discarded,
as these configurations typically correspond to high-energy states with large forces.
Including such structures would deteriorate the accuracy of the MLP fit.
Finally, we conducted high-throughput converged DFT calculations on the generated structures, discarding any that failed to converge.

\subsection{Validation of the SCS protocol}\label{subsec:validation}

\begin{figure}
	\centering
	\includegraphics[width=0.45\textwidth,trim = {0.0cm 0.0cm 0.0cm 0.0cm}, clip]{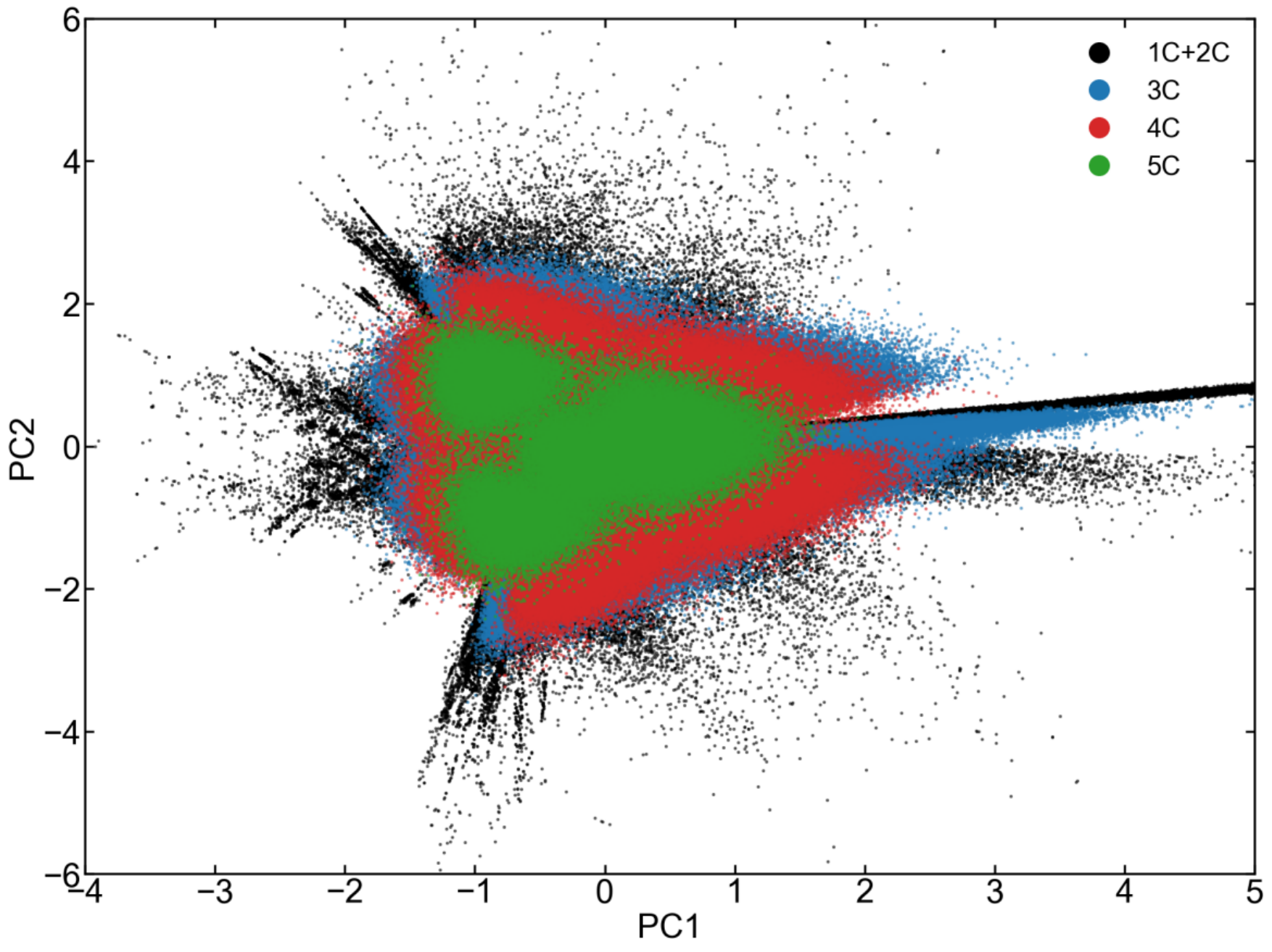}
\caption{Principal component analysis of structures with one and two elements (denoted as 1C+2C),
three elements (3C), four elements (4C), and five elements (5C).}
	\label{PCA}
\end{figure}

To assess the efficacy of the SCS protocol,
we performed systematic and careful validations on the TiZrHfCuNi
MPEA---an interesting system exhibiting continuous polyamorphic transition in its glass state~\cite{Cao2024}.
We first demonstrate that the SCS protocol can effectively generate a diverse and comprehensive training dataset.
To this aim, we performed principal component analysis (PCA) using the smooth overlap of atomic positions descriptor~\cite{doi:10.1021/acs.accounts.0c00403}.
The results are shown in Fig.~\ref{PCA}, where ``1C+2C" denotes the phase space sampled by the SCS protocol (limited to single- and two-element structures),
while ``3C", ``4C", and "5C" represent phase spaces for structures with three, four, and five elements, respectively.
Those multi-element structures (3C+4C+5C) were sampled from hybrid Monte Carlo/molecular dynamics (MC/MD) simulations~\cite{Thompson2022}.
Specifically, for 3C and 4C systems, we constructed equimolar BCC-like 96-atom supercells, considering all possible combinations:
10 ternary systems (e.g., TiZrHf, ZrHfCu, HfCuNi, etc.) and 5 quaternary systems (TiZrHfCu, ZrHfCuNi, etc.).
For 5C systems, we used an equimolar 120-atom BCC-like supercell.
Each structure underwent a 2-ns hybrid MC/MD simulation, yielding 1,000 equidistant configurations per system.
In total, we obtained 10,000 (3C), 5,000 (4C), and 1,000 (5C) supercells.
Figure~\ref{PCA} reveals that while the multi-element structures (3C, 4C, and 5C) exhibit diversity,
they occupy only a subset of the phase space covered by 1C+2C sampling.
Supplemental Material Fig.~S1~\cite{SM} further presents the computed extrapolation grades for multi-element structures (3C+4C+5C)
using the MTP trained on small-cell 1C+2C structures, evaluated via Shapeev's generalized D-optimality criterion~\cite{MTP2017}.
All structures exhibit extrapolation grades below 2.5, confirming their placement within the MTP's safe extrapolation regime~\cite{MTP2017}.
Notably, some structures show extrapolation grades below 1.0, indicating interpolation~\cite{MTP2017}.
The PCA analysis along with the extrapolation analysis underscore the efficacy of SCS in generating a broad and representative training dataset.

\begin{table*}[!htbp]
	\centering
	\caption{Performance evaluation of MTPs trained on different training datasets composed solely of one- and two-element small-cell structures.
		All models were validated on the same validation dataset, with RMSEs reported for energy, force, and stress tensor predictions.
		The notation, e.g., ``4-8-12",  denotes the MTP trained on a dataset sampled exclusively from 4-, 8-, and 12-atom cells (excluding MP near convex hull structures),
		and ``4-8-12-mp" represents the MTP trained on the same cell sizes but augmented with structures near the convex hull from the MP.
The labels ``All", ``SS", and ``IC" denote the RMSEs evaluated on the entire validation set, the solid-solution subset, and the ordered intermetallic compound subset, respectively.
	}
	\begin{ruledtabular}
		\begin{tabular}{cccccccccccc}
			& \multicolumn{2}{c}{Training dataset} & \multicolumn{3}{c}{Energy (meV/atom)} & \multicolumn{3}{c}{Force (meV/\AA)} & \multicolumn{3}{c}{Stress (GPa)}  \\
			\cline{2-3} \cline{4-6} \cline{7-9} \cline{10-12}
			& \#Structures & \#Atoms & IC & SS & All &  IC & SS & All &  IC & SS & All \\
			\hline
			4-8-12-16 & 32540 & 429124 & 33.7 &  16.3 & 19.1 & 107.4 & 129.7 & 128.5 & 1.14 & 0.63 & 0.71\\
			4-8-12-mp & 20346 & 242682 & 14.2 & 17.1  & 16.8 & 66.3 & 135.1  & 132.1 & 0.65 & 0.68  & 0.68\\
			4-8-12 & 19339 & 217908  & 33.1 & 17.5  & 19.9 & 116.9 & 140.9  & 139.6 & 1.29 & 0.76  & 0.81\\
			8-12 & 18807 & 215780  & 37.3 & 19.4  & 22.2 & 121.0 & 143.8  & 142.6 & 1.22 & 0.73  & 0.81\\
			4-12 & 16863 & 198100  & 44.3 & 20.1 & 24.2 & 126.4 & 156.3  & 154.7 & 1.32 & 0.87  & 0.93\\
			4-8 & 3008 & 21936  & 38.8 &  26.0 & 27.8 & 141.4 & 172.1 & 170.4 & 1.48 & 0.99  & 1.05\\
			12 & 16331 & 195972 & 42.4 & 21.1 & 24.5 & 125.9 & 152.2 & 150.8 &1.47 & 0.81  & 0.91\\
			8 & 2476 & 19808 & 39.5 & 26.2 & 28.1 & 131.8 & 167.2 & 165.4 & 1.46 & 0.94  & 1.01\\
			4 & 532 & 2128 & 218.9 & 53.5 & 89.6 &653.0 & 345.3 & 370.1 & 7.45& 2.45  & 3.42\\
		\end{tabular}
	\end{ruledtabular}
	\label{tab:performance}
\end{table*}

Next, we conducted systematic ablation experiments.
To ensure fair comparison, a consistent validation dataset containing diverse and comprehensive atomic environments was constructed.
Specifically, for each ternary configuration, we constructed 20 BCC-like supercells containing 90 atoms, as well as 20 FCC-like and 20 HCP-like supercells, each with 96 atoms.
Each supercell was designed with an equimolar composition, featuring a random distribution of different elements, followed by structural perturbations.
This process yielded a total of 600 structures across the 10 ternary systems.
A similar procedure was employed to generate quaternary validation structures, except that all supercells contained 96 atoms.
This process generated a total of 300 structures across the 5 quaternary systems.
For the quinary configurations, we constructed 40-atom BCC-like supercells and 80-atom FCC-like/HCP-like supercells.
Chemical compositions were sampled at 10\% intervals across the 10\% to 60\% elemental concentration range,
with lattice parameters dynamically adjusted based on the elemental compositions.
Following this, uniform scaling factors of 0.95, 1.00, and 1.05 were applied to the lattice parameters of each structure.
Additionally, Gaussian-distributed random displacements with a standard deviation of 0.15~$\AA$ were introduced to the atomic positions.
This process yielded a total of 1,134 quinary structures.
For the intermetallic compounds, we extracted binary structures situated within 0.04~eV/atom of the binary convex hull from the MP database.
Each selected structure was then subjected to a 2$\times$2$\times$2 or a 2$\times$2$\times$1 expansion, followed by similar lattice and atomic positions perturbations.
The final validation dataset comprises 2,298 structures in total, combining 2,034 solid-solution structures (ternary, quaternary, and quinary) with the 264 intermetallic compounds.
The ratio between solid-solution and intermetallic structures in the validation dataset closely mirrors that in the training dataset,
ensuring representative coverage across the multicomponent phase space of the TiZrHfCuNi system.

\begin{figure*}
	\centering
	\includegraphics[width=0.9\textwidth,trim = {0.0cm 0.0cm 0.0cm 0.0cm}, clip]{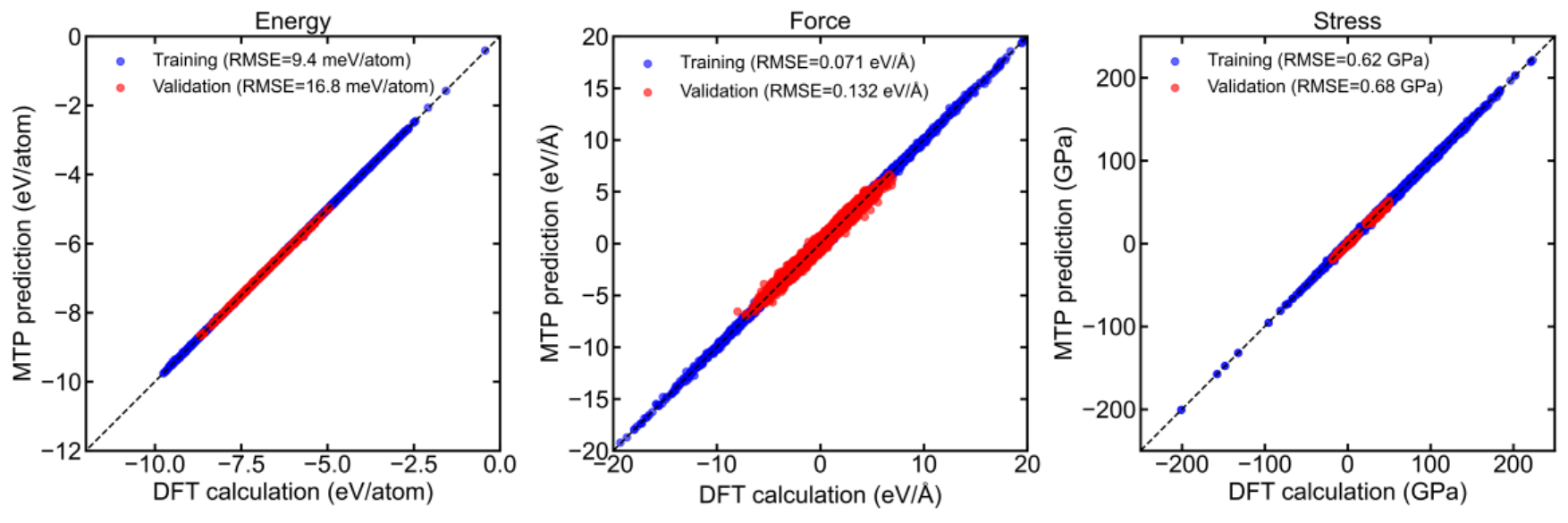}
	\caption{MTP predicted (a) energies, (b) forces, and (c) stress tensors against DFT results for the TiZrHfCuNi MPEA.
The MTP was trained using the SCS-generated dataset (i.e., the 4-8-12-mp dataset in Table~\ref{tab:performance}).
}
	\label{Fig3_TiZrHfCuNi_RMSE}
\end{figure*}

Based on the established validation dataset, we evaluated the performance of MTPs fitted using different training datasets.
We note that all training datasets contained only unary and binary small-cell structures, while the validation dataset included multicomponent (ternary, quaternary, and quinary)  large-cell structures.
The results of the ablation experiments are summarized in Table~\ref{tab:performance}.
We observe that for the solid-solution subset, sampling based solely on 4-atom or 8-atom cells---or their combination---proved ineffective, resulting in high validation RMSEs.
In contrast, using only 12-atom cells substantially improved MTP accuracy, which can be attributed to the significant increase in the number of atoms.
While supplementing the 12-atom set with either 4-atom or 8-atom cells alone did not yield further gains,
combining all three (4-, 8-, and 12-atom cells) led to a notable performance improvement over the 12-atom-only case.
This accuracy gain is particularly valuable given the only modest increase in computational cost.
Further extending the 4-8-12 training set by including sampled 16-atom cells led to a slight improvement in MTP accuracy.
However, this gain was marginal and accompanied by a sharp rise in computational cost---note that the 16-atom data in
Table~\ref{tab:performance} represent only one-sixth of the total structures generated by the enumeration method.
Although incorporating structures near the binary convex hull from the MP database (containing 1,007 structures and 24,774 atoms)
into the 4-8-12 set only marginally improved the model's performance on solid-solution structures,
it significantly reduced RMSEs for the intermetallic compound subset, as anticipated.
Overall, our results indicate that the MTP trained on the dataset constructed
via the SCS strategy---specifically the 4-8-12-mp dataset---achieves an optimal balance between accuracy and computational efficiency.
Figure~\ref{Fig3_TiZrHfCuNi_RMSE} provides a parity plot comparing the predictions of the 4-8-12-mp MTP against DFT reference values for both the training and validation datasets.

We also conducted an additional ablation study focusing on structural perturbations, specifically lattice and atomic position perturbations,
as well as different local atomic environments (BCC-, FCC-, and HCP-like).
The results of the ablation study are summarized in Supplemental Material Table S1 and Table S2~\cite{SM}.
Our findings indicate that incorporating atomic position perturbations leads to a significant improvement in force accuracy.
Additionally, the inclusion of lattice perturbations consistently reduces RMSEs for energies, forces, and stress tensors.
This suggests that lattice distortions can induce more substantial changes in atomic environments compared to atomic position perturbations alone.
The combination of both atomic position and lattice perturbations yields the best performance, underscoring their importance in creating a diverse and representative dataset.
Furthermore, the results from the ablation study on different local atomic environments reveal that FCC- and HCP-like cells encompass a broader range of atomic environments
compared to BCC-like cells. By combining all BCC-, FCC-, and HCP-like small cells,
one can sample a wider variety of atomic environments, which ultimately enhances the performance of the resulting MTP.

Going beyond evaluating the RMSEs, we further validated the SCS method by predicting physical properties using the 4-8-12-mp MLP.
The MLP predictions of energy-volume curves for unary and binary systems, compared to DFT results, are presented in Supplemental Material Fig.~S2~\cite{SM}.
For three-, four-, and five-element systems, the corresponding data can be found in Supplemental Material Fig.~S3~\cite{SM}.
For unary systems, we selected thermodynamically stable phases: HCP for Ti, Zr, and Hf, and FCC for Cu and Ni.
In contrast, the structures for multi-element systems were generated randomly, varying in unit cell types, sizes, and compositions.
Both the MACE and MTP models, trained on the SCS-generated small cells containing one or two elements,
effectively captured the energy-volume curves of multi-element systems, demonstrating overall good agreement with DFT results.
Notably, the MACE model outperformed the MTP model in accuracy (see Supplemental Material Fig.~S3~\cite{SM}), underscoring the benefits of equivariance in enhancing model performance.

We also predicted the phonon dispersion relations for structures on the binary convex hull.
The phonon dispersion relation, associated with the second derivative of the potential energy,
is particularly challenging to compute accurately, owing to its stringent requirement for high-precision force predictions.
As shown in Figs. S4 and S5 of the Supplemental Material~\cite{SM},
both the MACE and MTP models show good agreement with the reference DFT calculations,
with MACE again exhibiting superior accuracy.
For example, while the MTP model incorrectly predicts imaginary phonon modes in HfNi and TiNi, the MACE model correctly reproduces their phonon dispersion relations.
Our results also confirm that incorporating structures near the binary convex hull from the MP database is essential for accurately capturing phonon dispersions.
Interestingly, however, we observe that---except for HfNi, TiNi, and ZrNi---the 4-8-12 model without explicit intermetallic sampling still yields reasonably accurate phonon dispersions.
This suggests that the 4-8-12 dataset already captures a large portion of the relevant intermetallic phase space, albeit incompletely.

As a final test, we computed the temperature-dependent pair distribution functions using the MTP and compared them with experimental data~\cite{Cao2024}.
Except for the background not deducted from the experimental data,
we achieved overall good agreement between the MTP predictions and the experimental results (see Supplemental Material Fig.~S6~\cite{SM}).
All above tests collectively validate the efficacy of the SCS protocol in generating a diverse and representative training dataset for multicomponent systems,
effectively bypassing the computational overhead associated with iterative active learning cycles and large-cell DFT calculations.
In the following sections, we will investigate phase transformations in the TiZrVMo MPEA, the evolution of SRO parameters in the CoCrFeMnNi MPEA,
and the energy convex hulls and temperature-evolution SRO parameters in the AlTiZrNbHfTa MPEA, utilizing the MLPs developed through the SCS protocol.

\subsection{Phase transformation in the TiZrVMo MPEA}\label{subsec:TiZrVMo test}

\begin{figure}
	\begin{center}
		\includegraphics[width=0.48\textwidth,trim = {0.0cm 0.0cm 0.0cm 0.0cm}, clip]{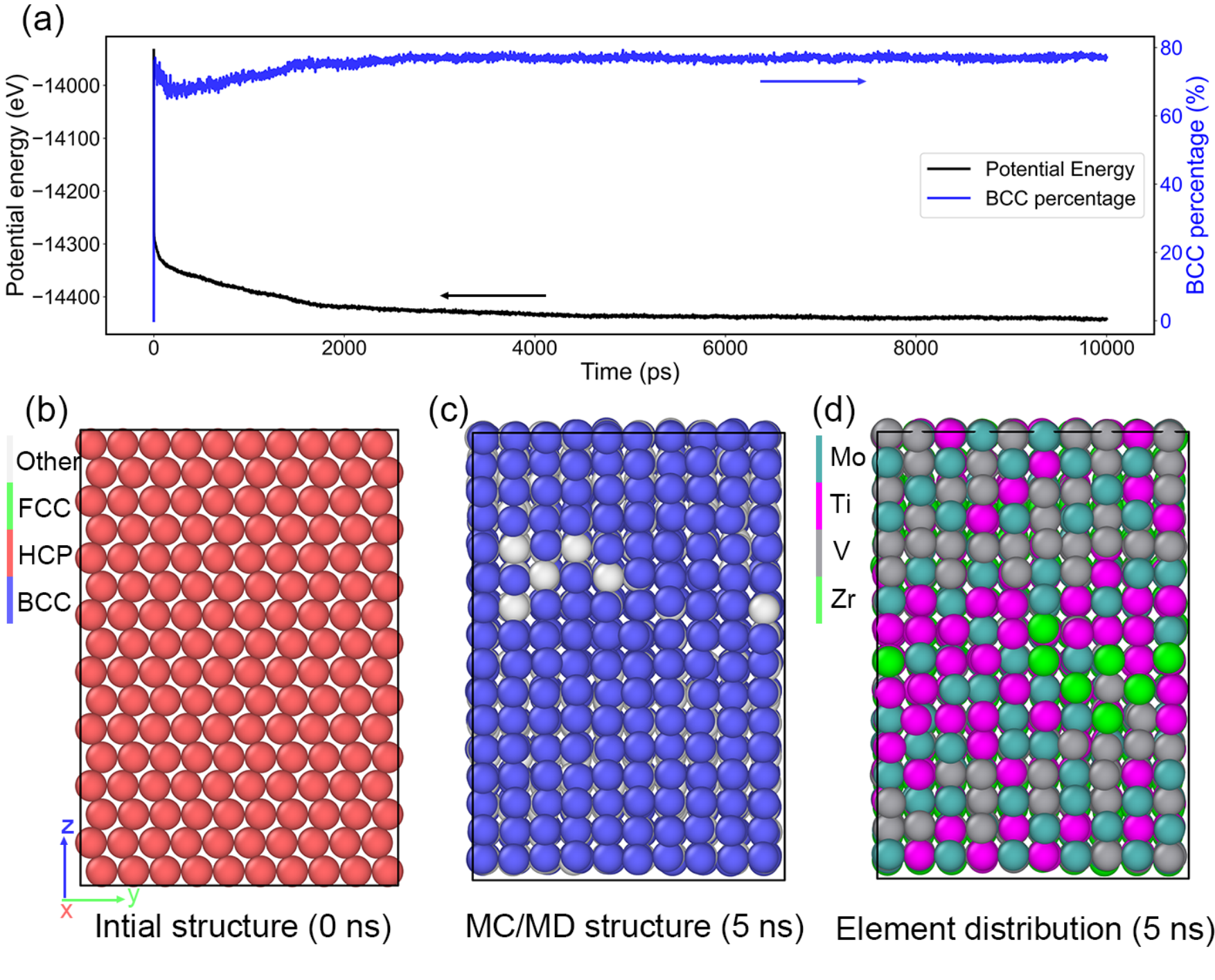}
	\end{center}
\caption{Transformation of the equimolar TiZrVMo MPEA from the initial HCP structure to the BCC structure during hybrid MC/MD simulations.
 (a) Initial configuration at 0 ns. (b)-(c) The structure after 5 ns and its elements distribution.}
	\label{Fig4_TiZrVMo_MCMD}
\end{figure}

TiZrVMo is a refractory high-entropy alloy, with experiments indicating a single-phase solid solution structure,
predominantly featuring a BCC structure with a small amount of HCP phase~\cite{MU2017668}.
To replicate this experimental observation, we trained a MTP specifically for this system using a dataset generated through the proposed SCS protocol.
To validate the accuracy of the trained MTP, we then performed hybrid MC/MD simulations for 5 ns at room temperature, starting from a 100-atom equimolar HCP-like supercell.
From the last 2 ns of the trajectory, we uniformly sampled 251 configurations to create our validation dataset.
The resulting validation RMSEs for energy, force, and stress were 1.9 meV/atom, 176 meV/$\AA$, and 0.23 GPa, respectively (see Supplemental Material Fig.~S7~\cite{SM}),
indicating a reasonable level of accuracy for the MTP potential.

Using the developed MTP potential, we conducted large-scale hybrid MC/MD simulations at room temperature, starting from a 1600-atom HCP-like supercell.
We employed Metropolis Monte Carlo parameters similar to those used by Song \emph{et al.}~\cite{NEP16-Song2024},
specifically 2000 MC atom exchange attempts after every 1000 MD steps.
The system reached equilibrium after  $10^{6}$ MD steps, at which the MC acceptance rate approached zero.
As illustrated in Fig.~\ref{Fig4_TiZrVMo_MCMD}(a), the HCP phase proved to be unstable, rapidly transforming into the BCC phase within a few hundred picoseconds.
This transformation is evidenced by significant changes in the potential energy and the percentage of the BCC phase present.
Our simulations successfully reproduced the findings of Song \emph{et al.}~\cite{NEP16-Song2024} [see Figs.~\ref{Fig4_TiZrVMo_MCMD}(b)-(d)]
and aligned well with experimental observations~\cite{MU2017668}.
The percentage of the transformed BCC phase approached approximately 80\% around the 3 ns and remained largely stable up to 10 ns.
These results demonstrate that the MLP developed through the SCS protocol can effectively capture phase transformations in MPEAs.

\subsection{SRO parameters in the CoCrFeMnNi MPEA}\label{subsec:CoCrFeMnNi test}

\begin{figure}
	\begin{center}
		\includegraphics[width=0.46\textwidth,trim = {0.0cm 0.0cm 0.0cm 0.0cm}, clip]{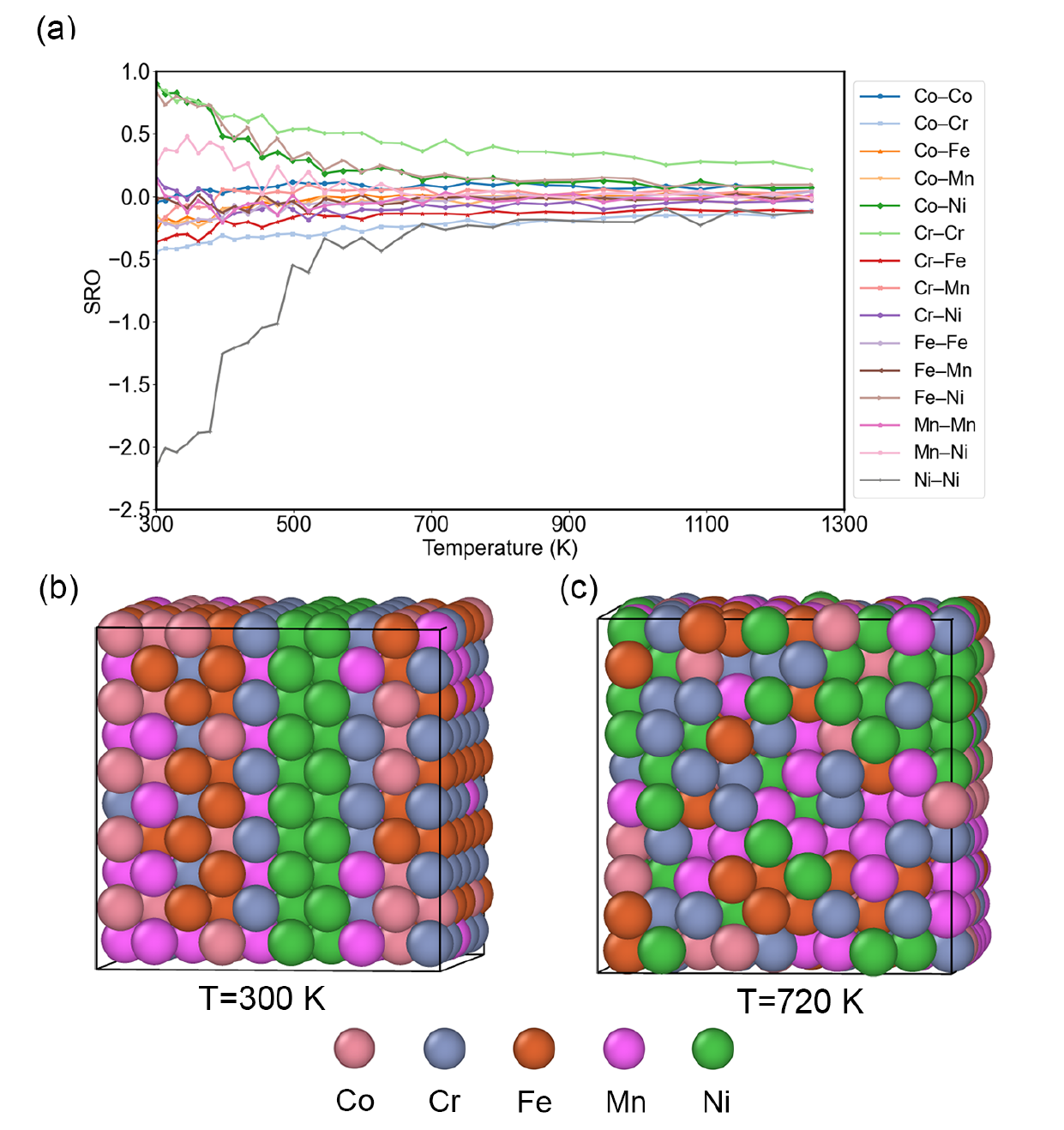}
	\end{center}
	\caption{(a) SRO parameters for the first shell in CoCrFeMnNi MPEA, averaged over the last 1000 steps and three independent runs.
		(b)-(c) Snapshots from hybrid MC/MD simulations at $T=300$ K and $T=720$ K, respectively.
		}
	\label{Fig5_CoCrFeMnNi_SRO}
\end{figure}

\begin{figure*}
	\begin{center}
		\includegraphics[width=0.9\textwidth,trim = {0.0cm 0.0cm 0.0cm 0.0cm}, clip]{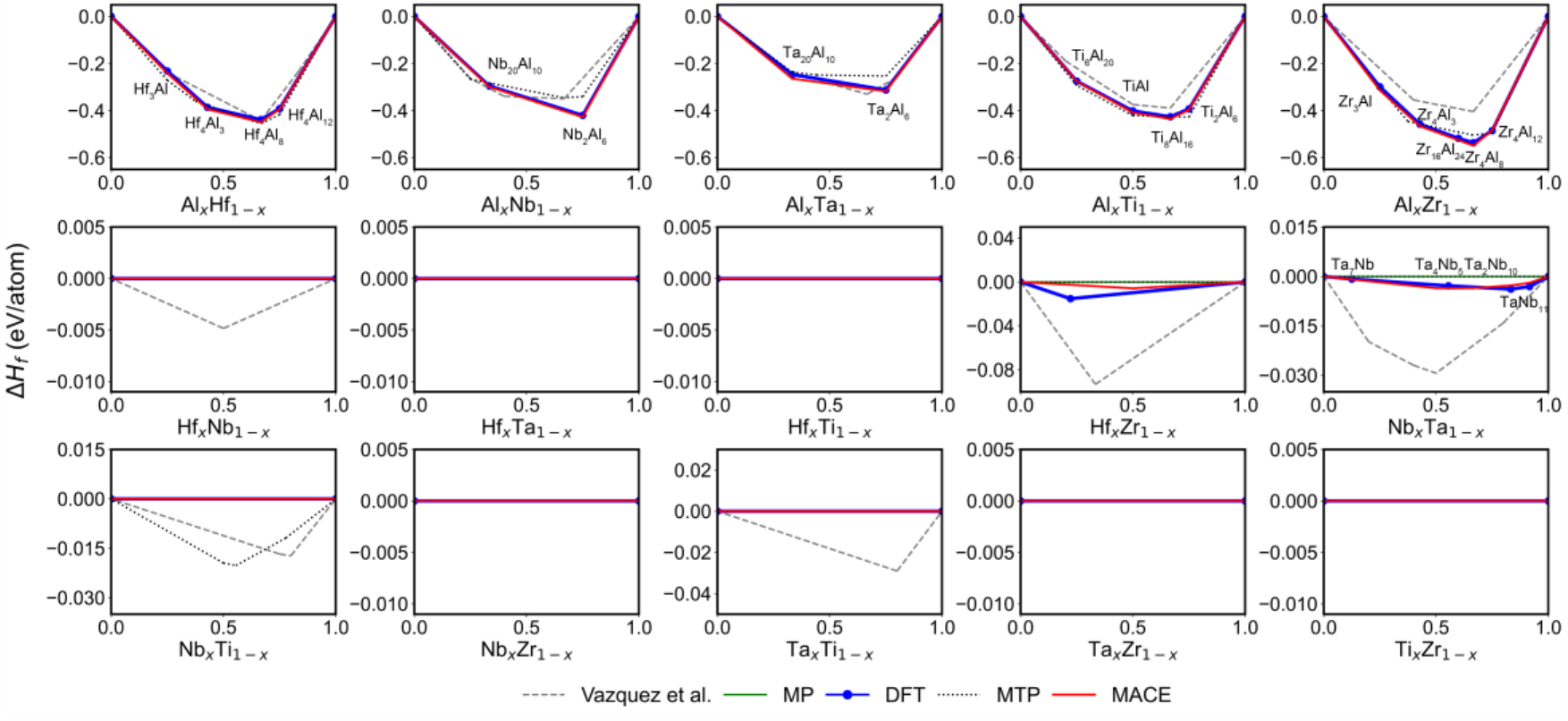}
	\end{center}
	\caption{Convex hull analysis for the AlTiZrNbHfTa MPEA by MTP, MACE, and DFT methods.
The convex hulls derived from Vazquez \textit{et al.}~\cite{Vazquez2024} are indicated by the dashed lines.
Note that only the data on the convex hull are shown.
	}
	\label{Fig6_AlTiZrNbHfTa_convex_hull}
\end{figure*}

As the second benchmark system, we evaluated the prototypical CoCrFeMnNi Cantor alloy~\cite{Cantor2004}.
The training dataset was generated using the SCS protocol,
and the resulting MTP was validated against a test dataset sampled from hybrid MC/MD simulations.
The initial configuration consisted of a 100-atom BCC-like supercell with equimolar composition.
Simulations were performed at 32 logarithmically spaced temperatures between 300 K and 1253 K,
each run for 2 ns, with 15 configurations extracted per temperature (totaling 480 test structures).
The validation RMSEs for energy, force, and stress were 7.0 meV/atom, 144 meV/$\AA$, and 0.26 GPa, respectively (Supplemental Material Fig.~S8~\cite{SM}),
indicating good accuracy of the MTP.

Using the fitted MTP, we computed the temperature-dependent Cowley's SRO parameters~\cite{PhysRev.77.669,Sheriff2024} through hybrid MC/MD simulations.
The Cowley's SRO parameters are extensively employed in the study of MPEAs to characterize the extent of short-range order.
Specifically, a value of 0 indicates fully disordered solid solutions, negative values suggest a tendency for the species to clustering, and a value of 1 signifies full separation.
For the hybrid MC/MD simulations,
we utilized a 500-atom FCC supercell, mirroring the configuration employed by Lopanitsyna \textit{et al.}~\cite{10.1103/PhysRevMaterials.7.045802},
and conducted 100 MC atom-exchange attempts following every 1000 MD steps.
This effectively reduces error accumulation from inadequate relaxation while ensuring convergence to stable states.
Each simulation at different temperatures lasted for 4ns and the configurations from the final 1ns were employed for computing the Cowley's SROs.

The calculated Cowley's SRO parameters are illustrated in Fig.~\ref{Fig5_CoCrFeMnNi_SRO}(a).
At low temperatures, we observe a strong attraction between Ni-Ni pairs, while Cr-Cr, Fe-Ni, and Co-Ni pairs exhibit significant repulsion.
As the temperature increases, the high-temperature phase gradually transitions toward a disordered state, as shown in Fig.~\ref{Fig5_CoCrFeMnNi_SRO}(c).
This behavior aligns with the findings of Lopanitsyna \emph{et al.}~\cite{10.1103/PhysRevMaterials.7.045802}.
The order-disorder phase transition occurs around 500 K, manifested by abrupt changes in the Ni-Ni SRO parameters.
Interestingly, a snapshot from the hybrid MC/MD simulations at 300 K reveals the formation of a Ni-Ni bilayer.
This observation contrasts with the results reported by Lopanitsyna \emph{et al.}~\cite{10.1103/PhysRevMaterials.7.045802},
where the two Ni layers were found to be separated.
We calculated the total energies of both structures using the MTP and DFT,
finding that the low-temperature structure with the Ni-Ni bilayer is 1.2 eV lower in energy
than the configuration with separate Ni layers reported in Ref.~\cite{10.1103/PhysRevMaterials.7.045802}.
This indicates a higher thermodynamic stability for the Ni-Ni bilayer structure.
Importantly, the pronounced Ni segregation persists even in 2000-atom supercells,
suggesting that the tendency to form Ni-Ni bilayers at low temperatures is independent of system size.
Consistent with the findings of Lopanitsyna \emph{et al.}~\cite{10.1103/PhysRevMaterials.7.045802},
the Cr atoms exhibit an ordered arrangement, distributed on both sides of the Ni layers, forming a simple cubic pattern.
Notably, the observed Ni segregation aligns with previous simulation results~\cite{Chen2021}
and correlates with experimentally observed phase separation in the equimolar CoCrFeMnNi MPEA~\cite{Glienke2020}.

\subsection{Thermodynamics and SRO parameters in the AlTiZrNbHfTa MPEA}\label{subsec:AlTiZrNbHfTa test}

\begin{figure*}
	\begin{center}
		\includegraphics[width=0.9\textwidth,trim = {0.0cm 0.0cm 0.0cm 0.0cm}, clip]{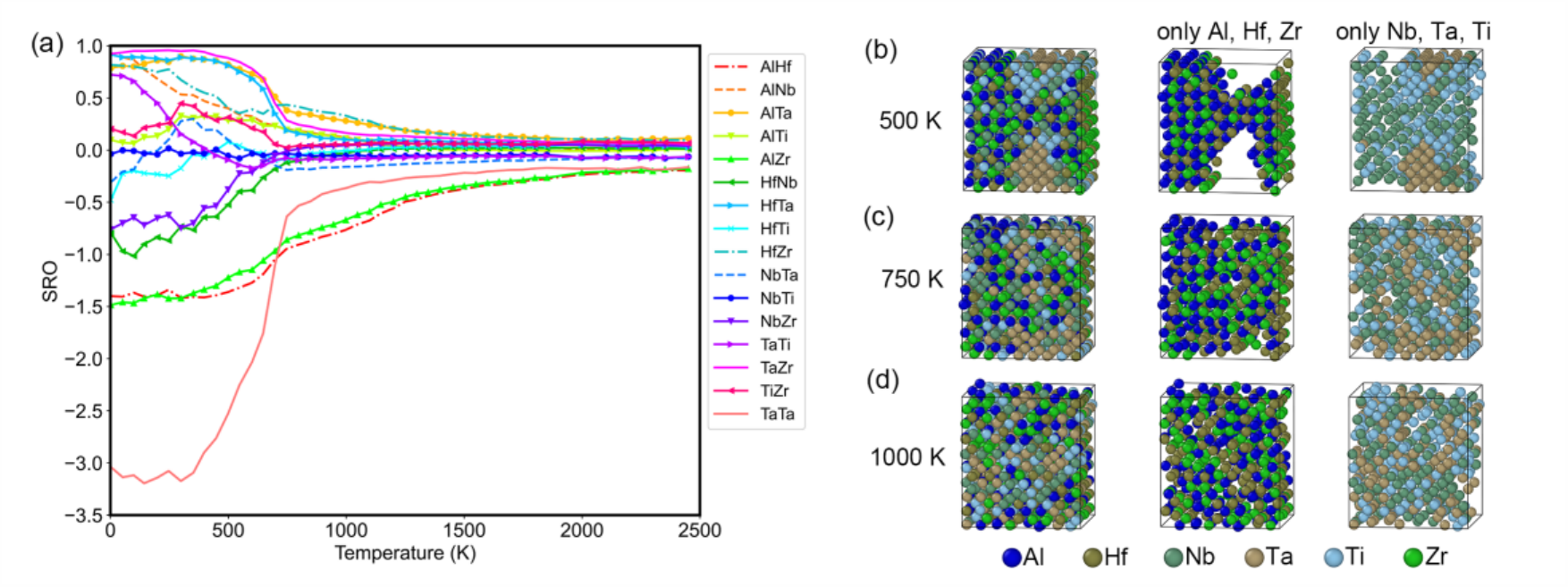}
	\end{center}
	\caption{SROs and snapshot structures for the AlTiZrNbHfTa MPEA.
(a) SROs as a function of temperature for the first-nearest-neighbor interactions.
 (b)-(d) Supercell structures from the last MC step at temperatures of 500 K, 750 K, and 1000 K, respectively.
 The first column displays all constituent atoms, the second column highlights only Al, Hf, and Zr atoms, while the third column focuses exclusively on Nb, Ta, and Ti atoms.
	}
	\label{Fig7_AlTiZrNbTaHf_SRO}
\end{figure*}

To further validate the generality of the SCS protocol for developing MLPs in MPEAs, we selected the refractory AlTiZrNbHfTa system as a benchmark. This MPEA system has attracted significant interest due to its promising combination of high strength, enhanced ductility, and low density~\cite{LIN2015100,SENKOV2014214}. Previous theoretical studies by Nataraj \emph{et al.} employed the CE method combined with DFT calculations to investigate its thermodynamic properties and phase stability~\cite{Nataraj2021_Acta,Nataraj2021}. More recently, Vazquez \emph{et al.} predicted the thermodynamics and chemical ordering of this system using the MLP-accelerated high-throughput CE approach~\cite{Vazquez2024}. These established references make this MPEA an ideal testbed for validating the efficacy of our SCS protocol.

Following the construction of the training dataset using the SCS protocol and subsequent MLP training, we generated a validation dataset through hybrid MC/MD simulations. These simulations employed a 54-atom BCC supercell with equimolar composition across 45 temperature points evenly spaced from 300~K to 2200~K. From the last 2~ns of production runs at each temperature, we collected 15 structures, yielding a total of 675 configurations for validation. The MLP predictions against DFT references showed RMSEs of 12.6~meV/atom for energy, 0.221~eV/$\AA$ for forces, and 0.42~GPa for stress tensors (see Supplemental Material Fig.~S9~\cite{SM}), demonstrating good accuracy. Notably, the MLP maintained high predictive fidelity even under significant thermal displacements, indicating that the SCS-developed potential effectively captures the high-temperature potential energy surface of this complex multi-element system.

We next evaluated the capability of SCS-developed MLPs to predict binary energy convex hulls---a challenging task
that requires precise treatment of configurational randomness and energy landscape optimization.
At each binary composition, we enumerated all possible configurations, performed structural relaxations using MLPs,
and designated the lowest-energy configuration as the ground-state structure.
Figure~\ref{Fig6_AlTiZrNbHfTa_convex_hull} compares the predicted convex hulls from MTP and MACE with those from DFT,
Vazquez \emph{et al.}~\cite{Vazquez2024} and the Materials Project database~\cite{MP_Jain2013}.
Both MACE and MTP models trained via the SCS protocol show good agreement with DFT references,
with MACE exhibiting superior accuracy attributable to its enhanced descriptive capacity.
This result indicates that SCS-developed MLPs can reliably describe not only solid solutions but also ordered intermetallic compounds.
Some discrepancy persists for the Hf-Zr system, likely due to insufficient sampling of intermetallic compounds that are incorrectly absent from the MP convex hull.
Note that the in-house trained M3GNet model by Vazquez \emph{et al.}~\cite{Vazquez2024} learned exclusively from BCC structures,
thereby being not able to accurately describe the global thermodynamic phase stability and leading to the largest deviations from the DFT references.

To further probe chemical ordering behavior, we computed temperature-dependent SRO parameters using MTP-driven hybrid MC/MD simulations. While employing the same initial BCC supercell as Vazquez \emph{et al.}~\cite{Vazquez2024}, we reduced the MC frequency to minimize relaxation errors, performing 200 MC atom-exchange attempts per 1000 MD steps over 4~ns simulations. The SRO parameters were computed from the final 1~ns trajectory using the Cowley's formulation~\cite{PhysRev.77.669,Sheriff2024}. As shown in Fig.~\ref{Fig7_AlTiZrNbTaHf_SRO}(a), all atomic pairs converge to zero SRO at high temperatures, indicating complete mixing. With decreasing temperature, distinct ordering patterns emerge: Al-Hf and Al-Zr pairs exhibit strong segregation tendencies, while Zr-Hf pairs show weak repulsion yet remain proximal to Al atoms. Notably, Ta-Ta pairs demonstrate significant separation that persists up to 750~K. Consistent with the findings of Vazquez \emph{et al.}~\cite{Vazquez2024}, at 500~K we observe co-segregation of Al, Hf, and Zr, while Nb, Ta, and Ti form separate clusters [Figs.~\ref{Fig7_AlTiZrNbTaHf_SRO}(b)-(d)]. This segregation diminishes at 750~K, with all elements mixing more uniformly, and homogenization further intensifies at 1000~K. Notably, our simulations indicate homogeneity at a lower temperature (750~K) compared to the CE studies by Nataraj \emph{et al.}~\cite{Nataraj2021_Acta,Nataraj2021} and Vazquez \emph{et al.}~\cite{Vazquez2024} (1000~K), a discrepancy likely attributable to vibrational entropy effects that are naturally incorporated in our MTP-MD simulations but absent in CE-assisted MC approaches.

\section{Conclusion}

In summary, we have developed an efficient SCS protocol
that enables the generation of diverse and representative training datasets for MPEAs
using only small-cell structures containing one or two elements.
The protocol systematically enumerates symmetry-inequivalent configurations across BCC-, FCC-, and HCP-like unit cells of 4, 8, and 12 atoms,
incorporates structures near binary convex hulls from the Materials Project database, and applies structural perturbations to enhance phase-space sampling.
Compared to conventional modeling approaches for MPEAs, this method substantially reduces computational costs
by eliminating the need for multiple iterative active learning cycles and expensive DFT calculations on supercells.
The efficacy of the SCS method has been rigorously validated
across four representative MPEA systems---TiZrHfCuNi, TiZrVMo, CoCrFeMnNi, and AlTiZrNbHfTa---demonstrating
accurate descriptions of chemical ordering, order-disorder phase transitions, and thermodynamic properties.
By establishing an efficient and transferable one-shot sampling protocol for constructing high-quality training datasets across multiple elements,
this work provides a solid foundation for developing universal MLPs for MPEAs.
For systems requiring enhanced accuracy, the SCS protocol can be combined with active learning methods
to sample critical configurations, further extending its applicability and predictive power.

\section*{Acknowledgements}
This work is supported by
the National Natural Science Foundation of China (Grants No.~52422112 and No.~52188101),
the Strategic Priority Research Program of the Chinese Academy of Sciences (XDA041040402),
the Liaoning Province Science and Technology Major Project (2024JH1/11700032, 2023021207-JH26/103 and RC230958),
the National Science and Technology Major Project (No. 2025ZD0618901)
and
the Special Projects of the Central Government in Guidance of Local Science and Technology Development (2024010859-JH6/1006).
P.L. would like to express gratitude to Prof. M. Ceriotti for generously sharing the structural data of the CoCrFeMnNi MPEA.

\section*{Author contributions}
P.L. and X.-Q.C. conceived and supervised the project.
P.L. and Y.L. designed the research.
Y.L. performed the calculations.
J.W., H.D., and Y.S. participated in discussions.
P.L. and Y.L. wrote the manuscript with inputs from other authors.
All authors comments on the manuscript.

\section*{Competing Interests}
The authors declare no competing interests.

\section*{Data Availability}
The source data that support the findings of this study will be publicly available upon acceptance of the paper.

\bibliography{Reference} 

\end{document}


\title{Supplemental Material to \\
``Efficient small-cell sampling for machine-learning potentials of multi-principal element alloys"}

\author{Yan Liu}
\affiliation{Shenyang National Laboratory for Materials Science, Institute of Metal Research, Chinese Academy of Sciences, 110016 Shenyang, China}
\affiliation{School of Materials Science and Engineering, University of Science and Technology of China, 110016 Shenyang, China}

\author{Jiantao Wang}
\affiliation{Shenyang National Laboratory for Materials Science, Institute of Metal Research, Chinese Academy of Sciences, 110016 Shenyang, China}
\affiliation{School of Materials Science and Engineering, University of Science and Technology of China, 110016 Shenyang, China}

\author{Hongkun Deng}
\affiliation{Shenyang National Laboratory for Materials Science, Institute of Metal Research, Chinese Academy of Sciences, 110016 Shenyang, China}
\affiliation{School of Materials Science and Engineering, University of Science and Technology of China, 110016 Shenyang, China}

\author{Yan Sun}
\affiliation{Shenyang National Laboratory for Materials Science, Institute of Metal Research, Chinese Academy of Sciences, 110016 Shenyang, China}

\author{Xing-Qiu Chen}
\email{xingqiu.chen@imr.ac.cn}
\affiliation{Shenyang National Laboratory for Materials Science, Institute of Metal Research, Chinese Academy of Sciences, 110016 Shenyang, China}

\author{Peitao Liu}
\email{ptliu@imr.ac.cn}
\affiliation{Shenyang National Laboratory for Materials Science, Institute of Metal Research, Chinese Academy of Sciences, 110016 Shenyang, China}

\maketitle

\begin{figure*}
	\centering
	\includegraphics[width=0.95\textwidth,trim = {0.0cm 0.0cm 0.0cm 0.0cm}, clip]{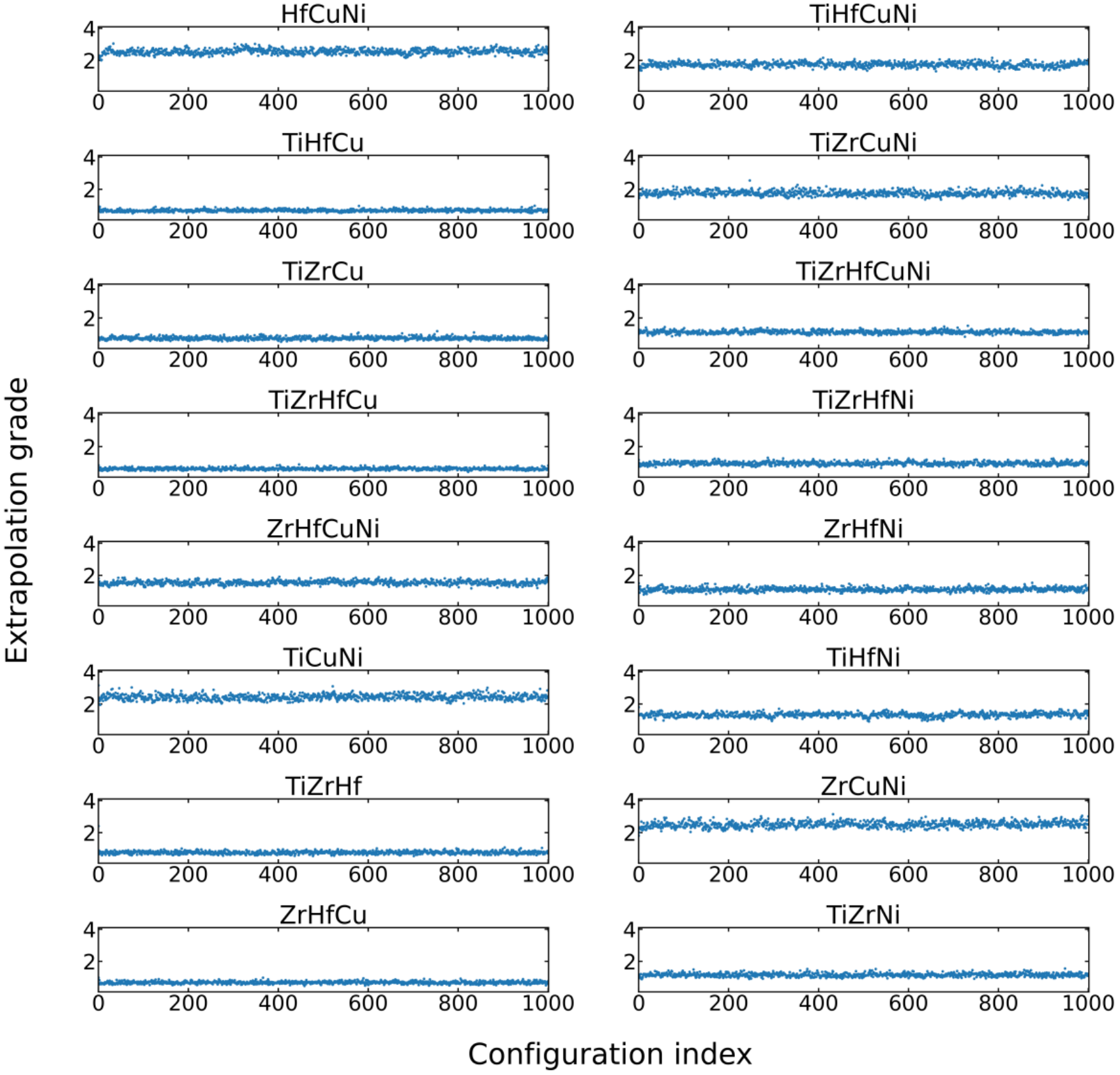}
	\caption{Extrapolation grades of the structures with three-, four-, and five elements, predicted by the MTP fitted with small-cell structures consisting of only one- and two-elements.}
	\label{FigS1_extrapolation_grades}
\end{figure*}

\begin{figure}
	\centering
	\includegraphics[width=0.95\textwidth,trim = {0.0cm 0.0cm 0.0cm 0.0cm}, clip]{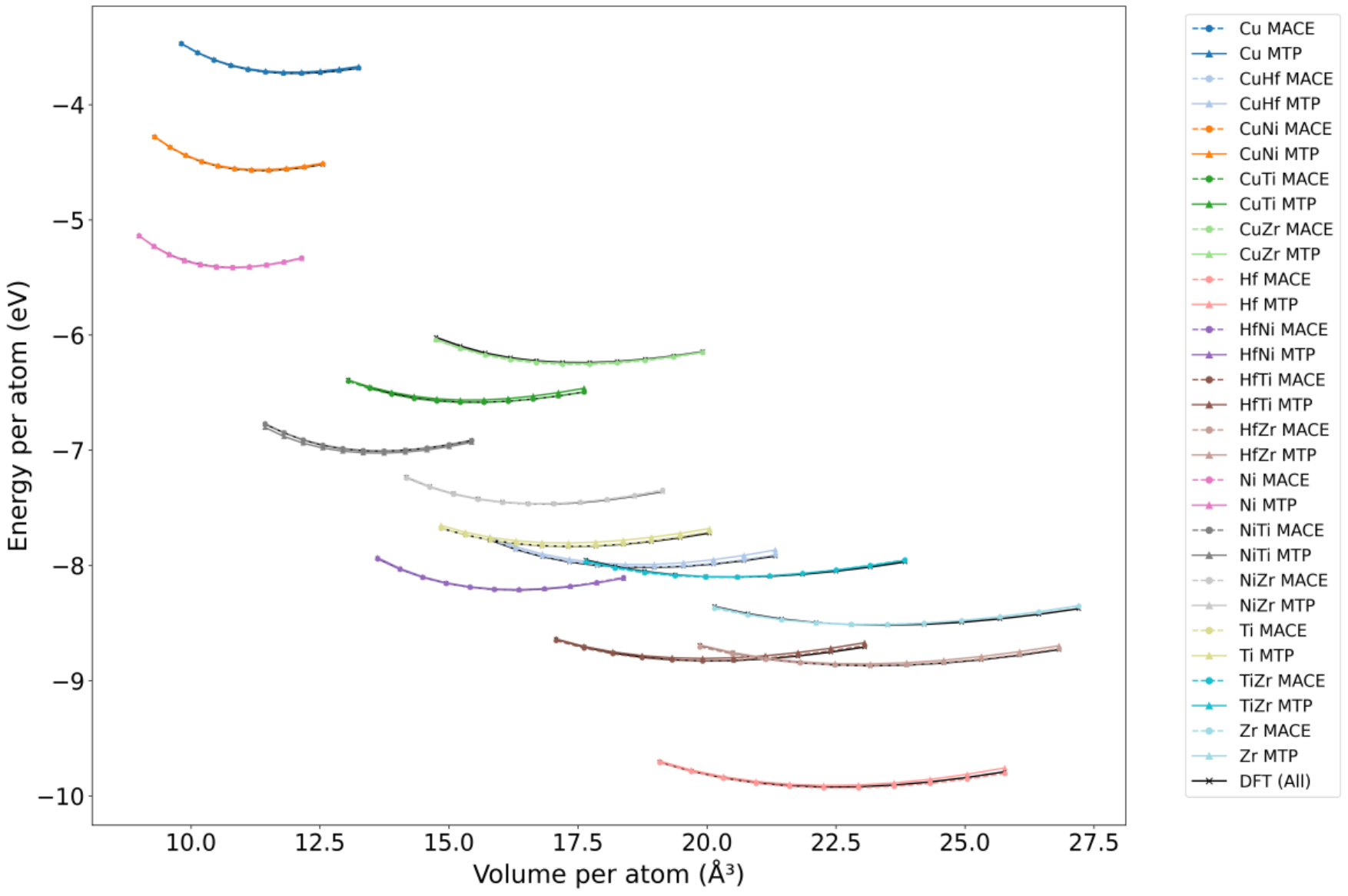}
	\caption{Energy-volume curves for unary and binary systems predicted by MACE, MTP, and DFT.
Note that the MACE and MTP  fitted with small-cell structures consisting of only one- and two-elements.}
	\label{FigS2_EV_Unary_Binary_Systems}
\end{figure}

\begin{figure}
	\centering
	\includegraphics[width=0.95\textwidth,trim = {0.0cm 0.0cm 0.0cm 0.0cm}, clip]{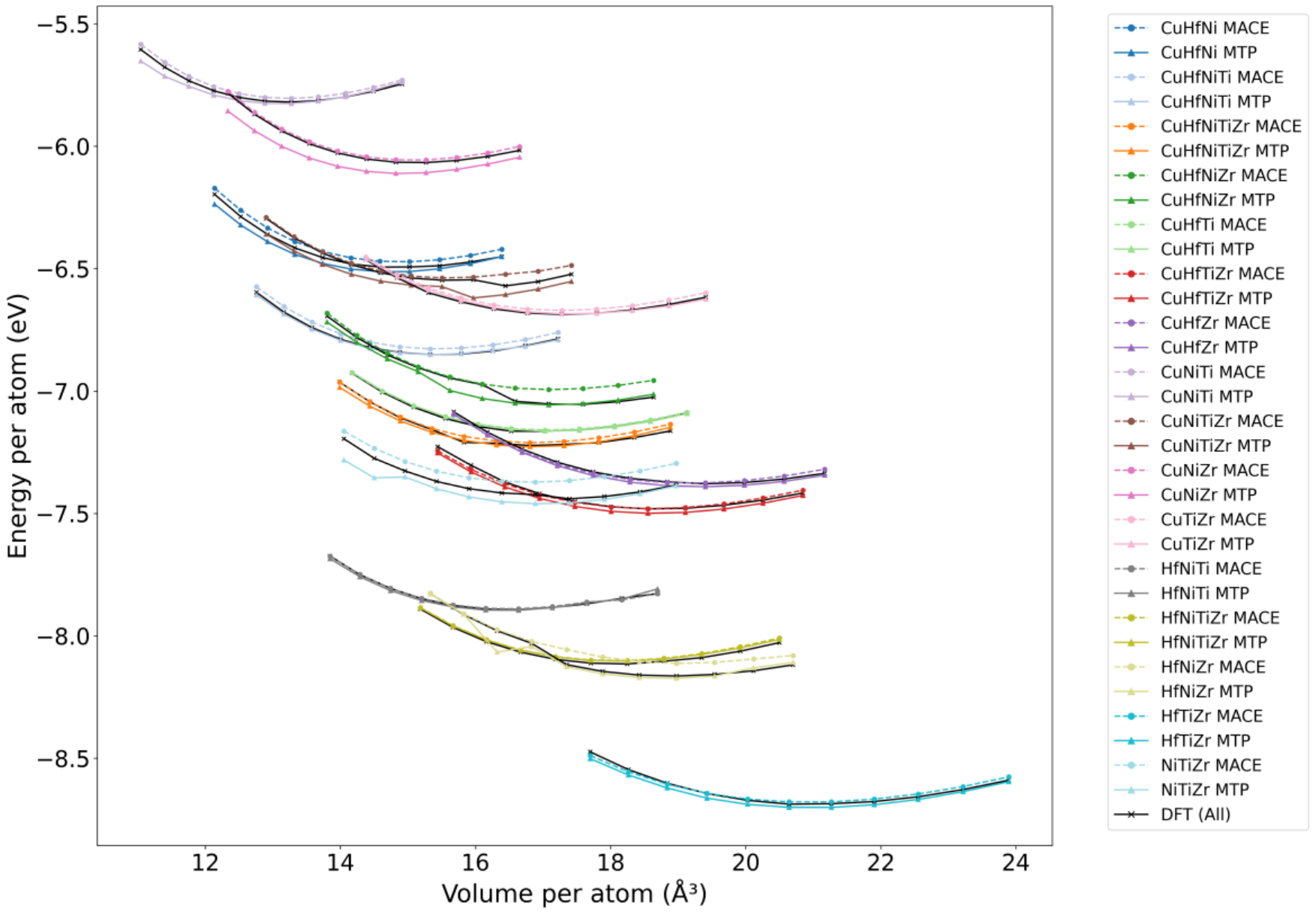}
	\caption{Energy-volume curves for three-, four-, and five-elements systems by MACE, MTP, and DFT.
Note that the MACE and MTP fitted with small-cell structures consisting of only one- and two-elements.}
	\label{FigS3_EV_3C_4C_5C_Systems}
\end{figure}

\begin{figure}
	\centering
	\includegraphics[width=0.95\textwidth,trim = {0.0cm 0.0cm 0.0cm 0.0cm}, clip]{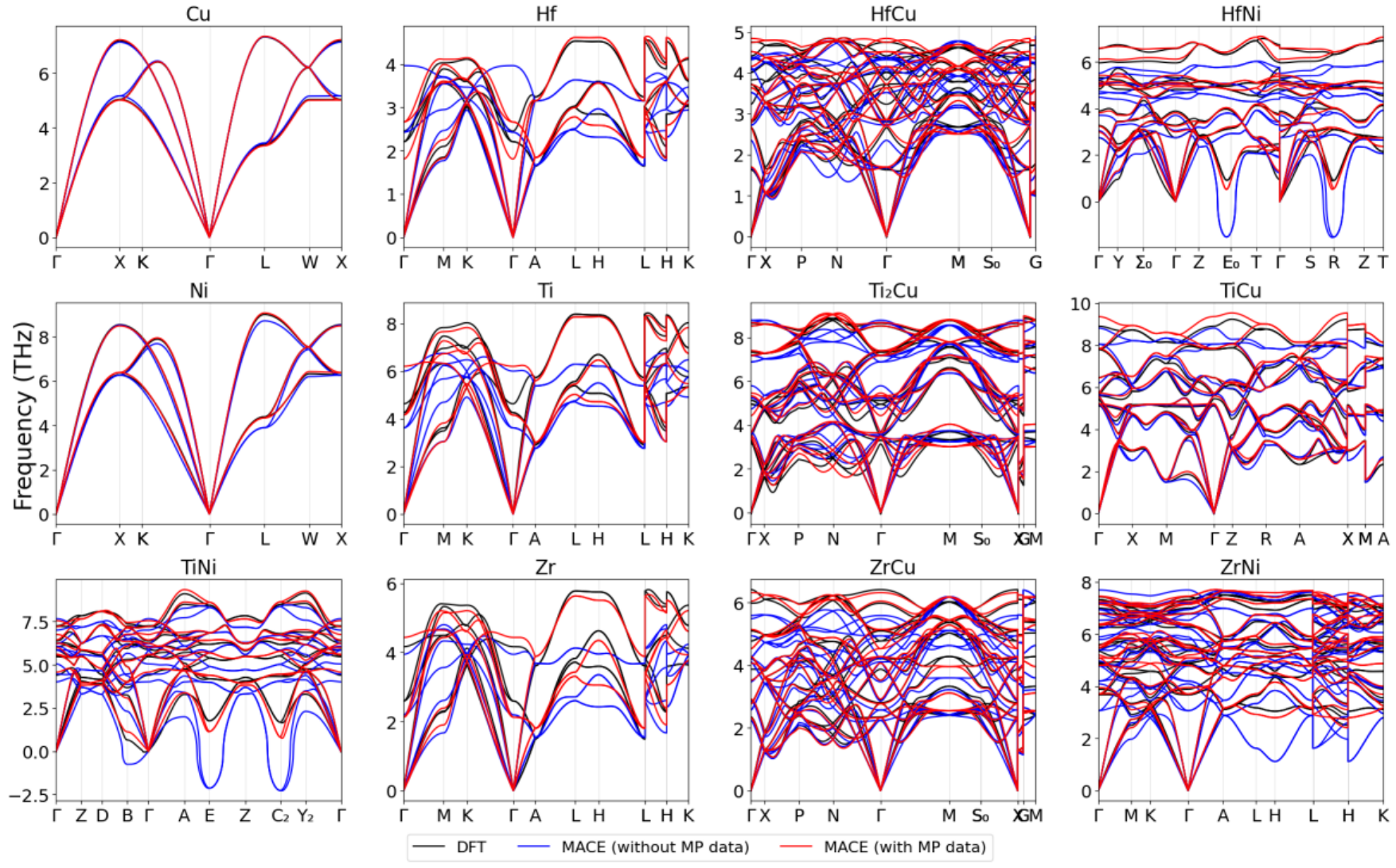}
	\caption{Phonon dispersion relationships for the structures on the binary energy convex hull, predicted by MACE and DFT.
}
	\label{FigS4_phonon_MACE_DFT}
\end{figure}

\begin{figure}
	\centering
	\includegraphics[width=0.95\textwidth,trim = {0.0cm 0.0cm 0.0cm 0.0cm}, clip]{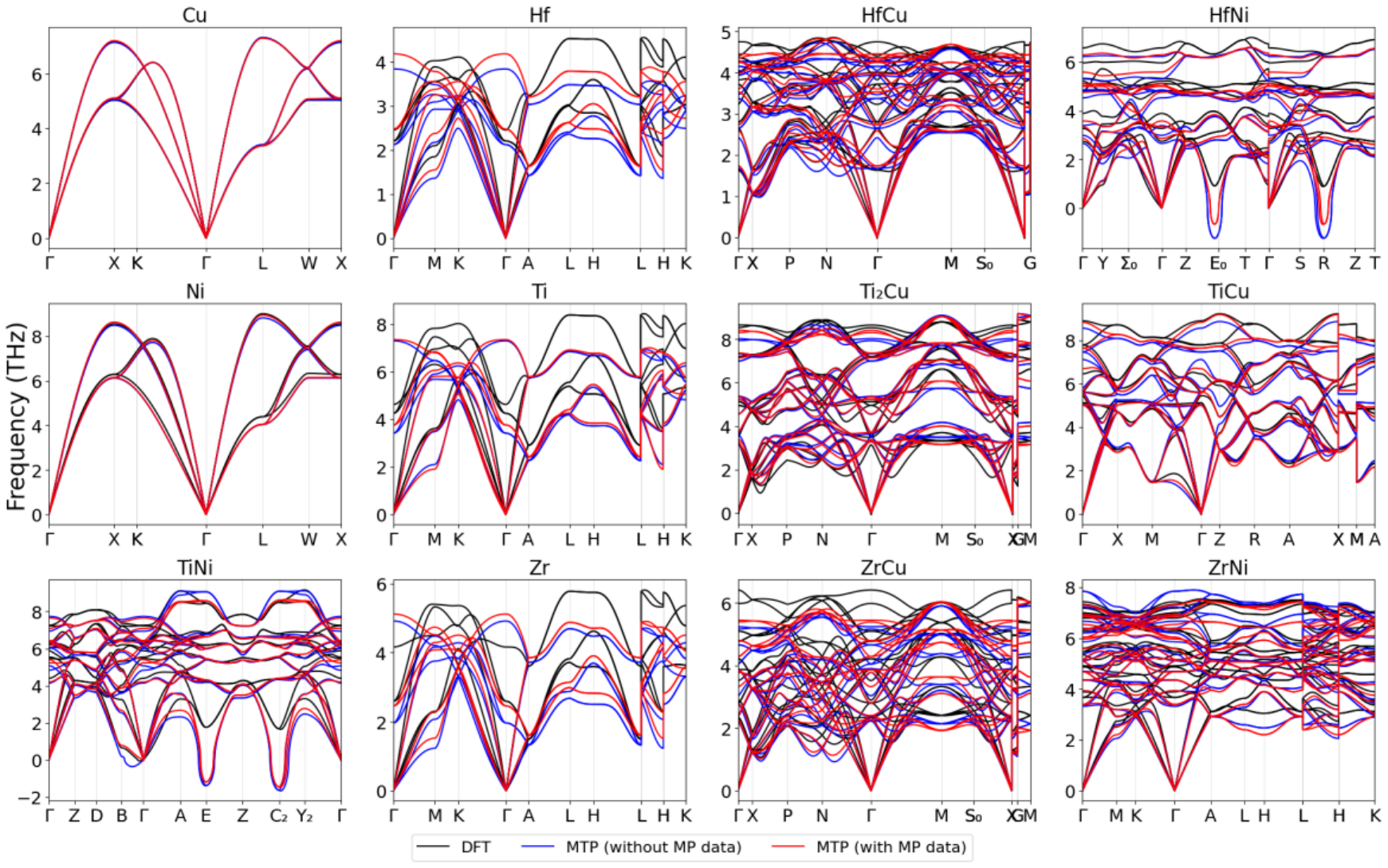}
	\caption{Phonon dispersion relationships for the structures on the binary energy convex hull, predicted by MTP and DFT.}
	\label{FigS5_phonon_MTP_DFT}
\end{figure}

\begin{figure}
	\centering
	\includegraphics[width=0.8\textwidth,trim = {0.0cm 0.0cm 0.0cm 0.0cm}, clip]{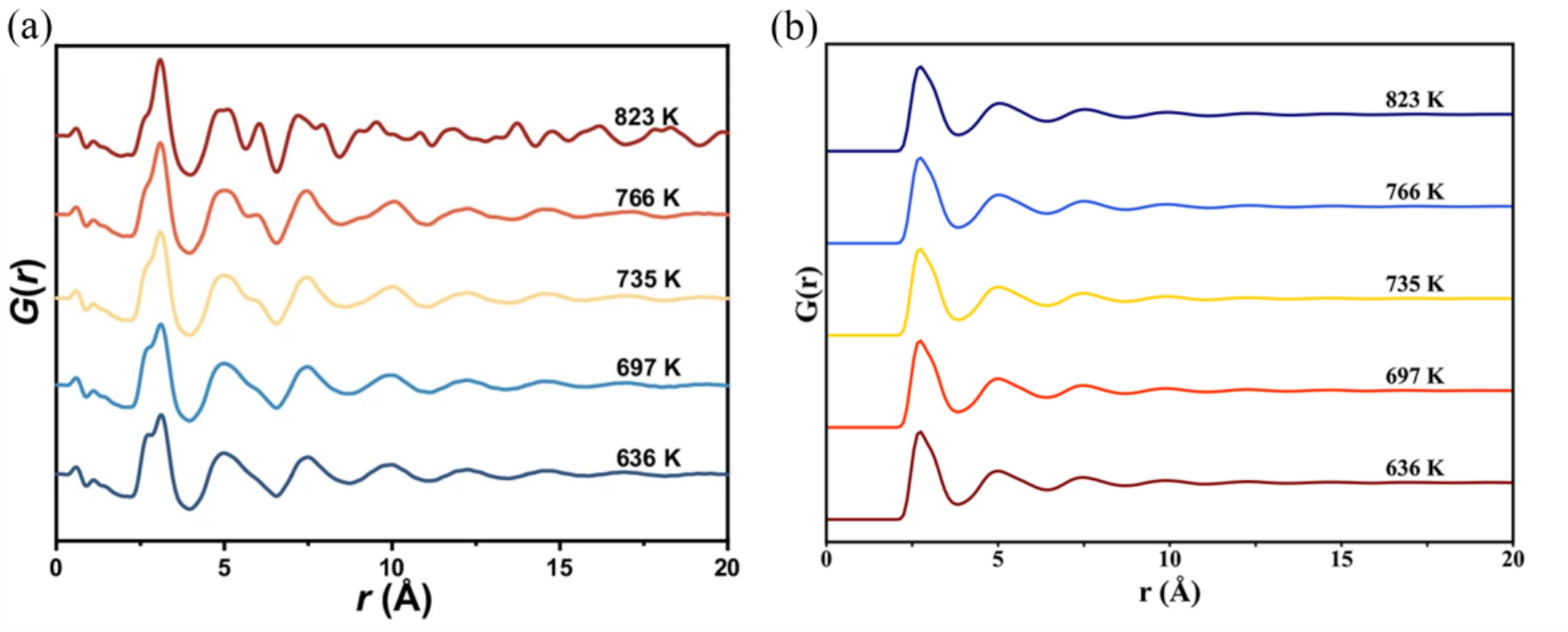}
	\caption{The experimentally measured (a) and MTP-predicted (b) pair distribution functions of the equimolar TiZrHfCuNi HEA at various temperatures.
The MC/MD simulations employed a 2000-atom supercell. The experimental data were taken from Ref.~\cite{Cao2024}.}
	\label{FigS6_TiZrHfCuNi_RDF}
\end{figure}

\begin{figure}
	\begin{center}
		\includegraphics[width=0.85\textwidth,trim = {0.0cm 0.0cm 0.0cm 0.0cm}, clip]{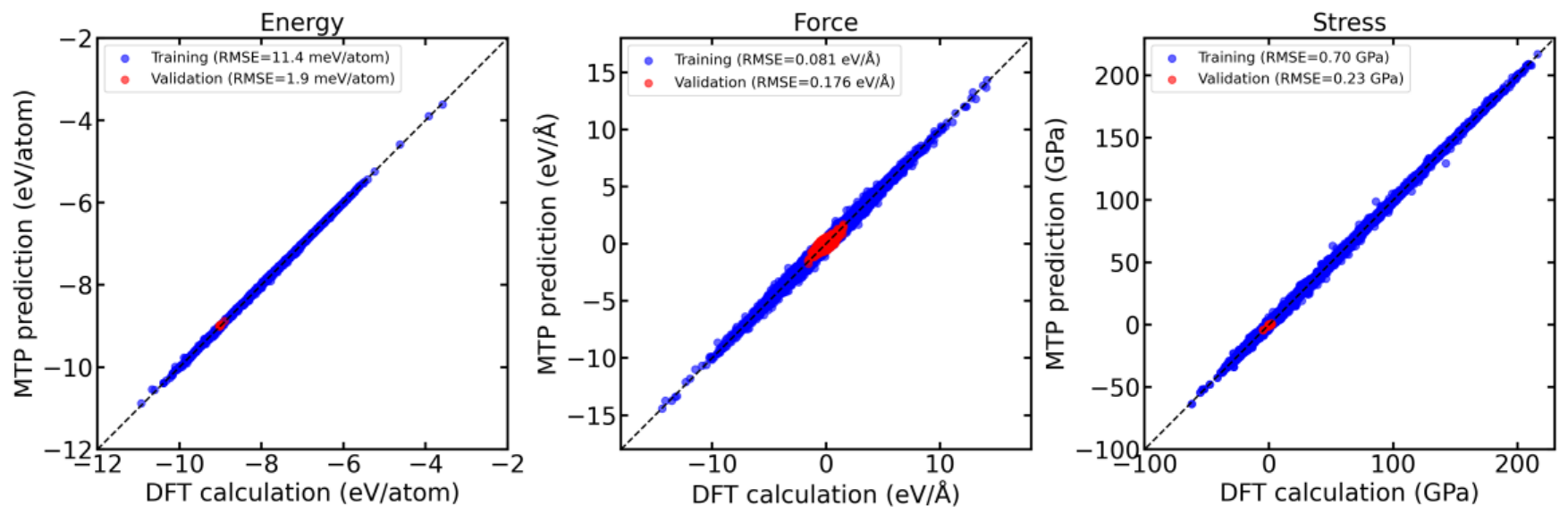}
	\end{center}
	\caption{MTP predicted (a) energies, (b) forces, and (c) stress tensors against DFT results for the TiZrVMo HEA.
	}
	\label{FigS7_TiZrVMo_RMSE}
\end{figure}

\begin{figure}
	\begin{center}
		\includegraphics[width=0.85\textwidth,trim = {0.0cm 0.0cm 0.0cm 0.0cm}, clip]{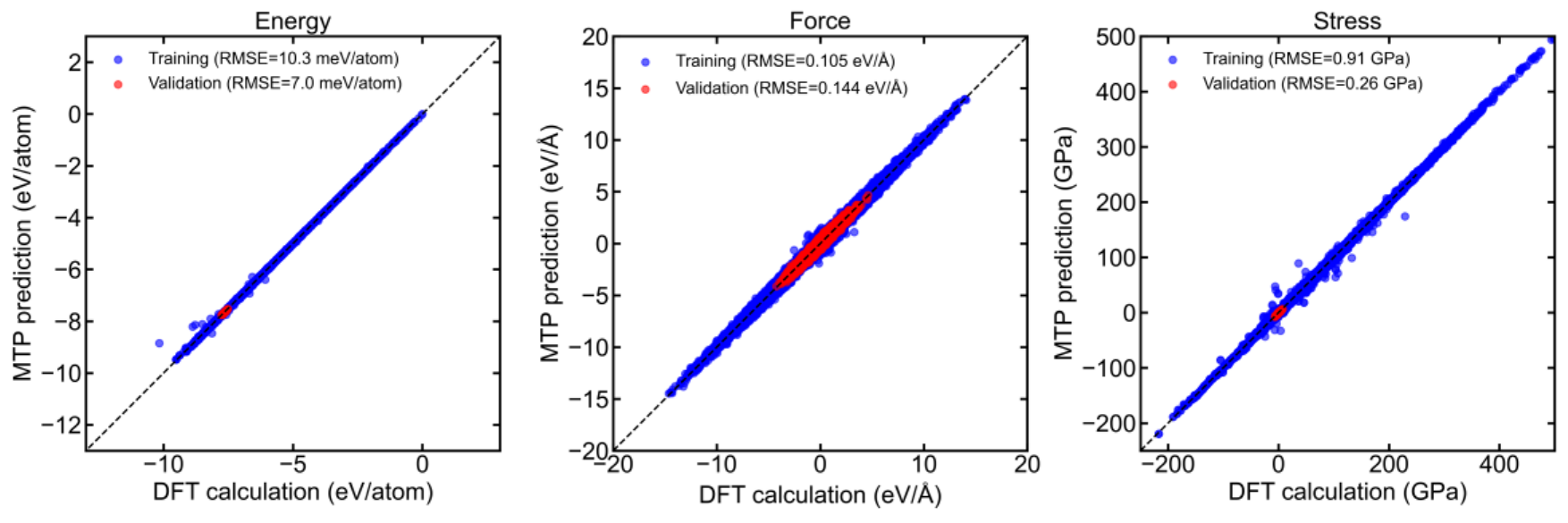}
	\end{center}
	\caption{MTP predicted (a) energies, (b) forces, and (c) stress tensors against DFT results for the CoCrFeMnNi HEA.
	}
	\label{FigS8_CoCrFeMnNi_RMSE}
\end{figure}

\begin{figure}
	\begin{center}
		\includegraphics[width=0.85\textwidth,trim = {0.0cm 0.0cm 0.0cm 0.0cm}, clip]{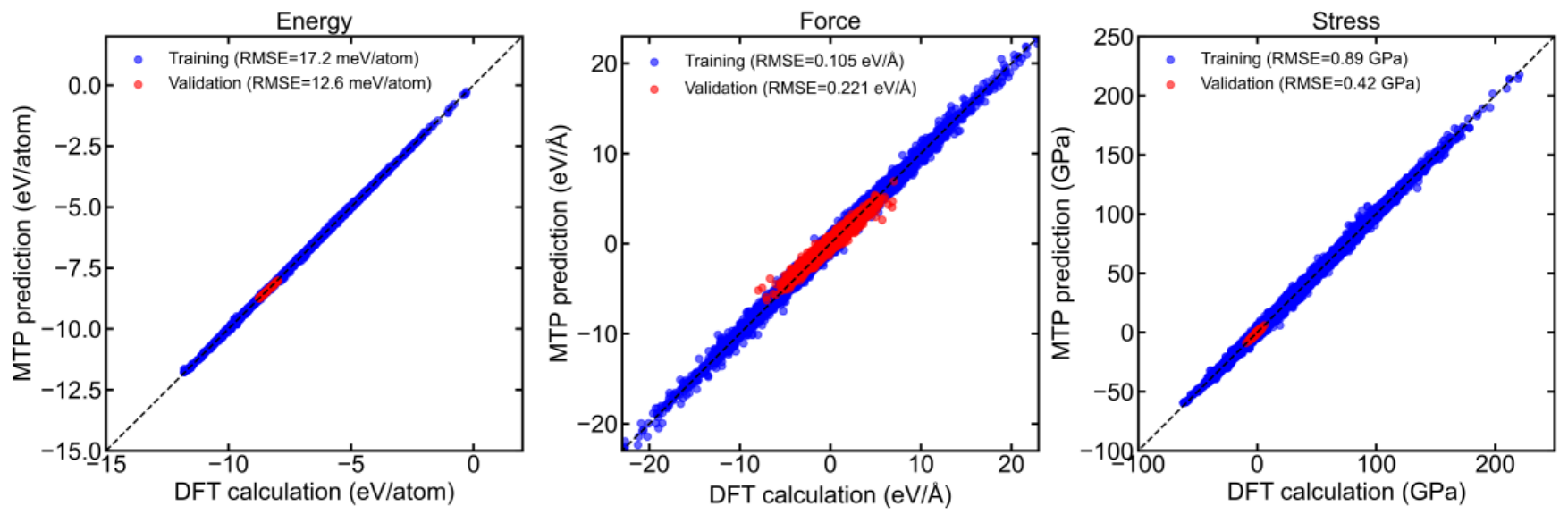}
	\end{center}
	\caption{MTP predicted (a) energies, (b) forces, and (c) stress tensors against DFT results for the AlTiZrNbHfTa HEA.
	}
	\label{FigS9_AlTiZrNbHfTa_RMSE}
\end{figure}

\clearpage

\begin{table*}[!htbp]
	\centering
	\caption{Impact of lattice and atomic position perturbations on the diversity of training structures used for fitting MTPs.}
	\begin{ruledtabular}
		\begin{tabular}{lccccc}
			& \multicolumn{2}{c}{Number} & Energy & Force & Stress \\
			\cline{2-3} \cline{4-4} \cline{5-5} \cline{6-6}
			& Structures & Atoms & (meV/atom) & (meV/\AA) & (GPa) \\
			\hline
			Both lattice and atomic position perturbations & 20346 & 242682 & 16.8 & 132.1 & 0.68 \\
			Only lattice perturbations & 19455 & 219720 & 22.4 & 153.5 & 0.96 \\
			Only atomic position perturbations & 6485 & 73240 & 32.4 & 176.2 & 1.52 \\
			No lattice and atomic position perturbations & 6485 & 73240 & 25.9 & 230.7 & 1.45 \\
		\end{tabular}
	\end{ruledtabular}
	\label{tab:performance_Perturbation}
\end{table*}

\begin{table}[!htbp]
	\centering
	\caption{Impact of different structural configurations on the diversity of training structures used for fitting MTPs.}
	\begin{ruledtabular}
		\begin{tabular}{lccccc}
			& \multicolumn{2}{c}{Number} & Energy & Force & Stress \\
			\cline{2-3} \cline{4-4} \cline{5-5} \cline{6-6}
			& Structures & Atoms & (meV/atom) & (meV/\AA) & (GPa) \\
			\hline
			BCC+FCC+HCP & 20346 & 242682 & 16.8 & 132.1 & 0.68 \\
			BCC+FCC     & 11160 & 125760 & 25.5 & 143.1 & 0.83 \\
			FCC+HCP     & 13920 & 157200 & 23.9 & 164.3 & 1.04 \\
			BCC+HCP     & 13830 & 156480 & 28.7 & 164.9 & 1.14 \\
			FCC         & 5625 & 63240 & 29.6 & 176.3 & 1.17 \\
			HCP         & 8295 & 93960 & 26.5 & 191.3 & 1.19 \\
			BCC         & 5535 & 62520 & 37.9 & 216.7 & 1.62 \\
		\end{tabular}
	\end{ruledtabular}
	\label{tab:performance_structures}
\end{table}

%

\bibliography{Reference} 